\documentstyle[epsfig]{mn2e}

\def\gtsima{$\; \buildrel > \over \sim \;$}
\def\ltsima{$\; \buildrel < \over \sim \;$}
\def\gtrsim{\lower.5ex\hbox{\gtsima}}
\def\lesssim{\lower.5ex\hbox{\ltsima}}

\begin{document}

\title[BHs and metallicity - I. X-ray binaries]{Dynamics of stellar black holes in young star clusters with different metallicities - I. Implications for X-ray binaries}
\author[Mapelli et al.]
{M. Mapelli$^{1}$, L. Zampieri$^{1}$, E. Ripamonti$^{2}$, A. Bressan$^{3}$
\\
$^1$INAF-Osservatorio Astronomico di Padova, Vicolo dell'Osservatorio 5, I--35122, Padova, Italy; {\tt michela.mapelli@oapd.inaf.it}\\
$^2$Universit\`a Milano Bicocca, Dipartimento di Fisica G. Occhialini, Piazza della Scienza 3, I--20126, Milano, Italy\\
$^3$Scuola Internazionale Superiore di Studi Avanzati (SISSA), Via Bonomea 265, I--34136, Trieste, Italy
}
\maketitle \vspace {7cm }

  \begin{abstract}
We present  $N-$body simulations of intermediate-mass ($3000-4000$ M$_\odot{}$) young star clusters (SCs) with three different metallicities ($Z=0.01$, 0.1 and 1 Z$_\odot{}$), including metal-dependent stellar evolution recipes and binary evolution. Following recent theoretical models of wind mass loss and core collapse supernovae, we assume that the mass of the stellar remnants depends on the metallicity of the progenitor stars. In particular, massive metal-poor stars ($Z\le{}0.3$ Z$_\odot{}$) are enabled to form massive stellar black holes (MSBHs, with mass $\ge{}25$ M$_\odot{}$) through direct collapse. We find that three-body encounters, and especially dynamical exchanges, dominate the evolution of the MSBHs formed in our simulations. In SCs with $Z=0.01$ and 0.1 Z$_\odot{}$, about 75 per cent of simulated MSBHs form from single stars and become members of binaries through dynamical exchanges in the first 100 Myr of the SC life. This is a factor of $\gtrsim{}3$ more efficient than in the case of low-mass ($<25$ M$_\odot{}$) stellar black holes. A small but non-negligible fraction of MSBHs power wind-accreting ($10-20$ per cent) and Roche lobe overflow (RLO, $5-10$ per cent) binary systems. The vast majority of MSBH binaries that undergo wind accretion and/or RLO were born from dynamical exchange. This result indicates that MSBHs can power X-ray binaries in low-metallicity young SCs, and is very promising to explain the association of many ultraluminous X-ray sources with low-metallicity and actively star forming environments.
\end{abstract}

%

\section{Introduction}

The mass spectrum of black holes (BHs) that form from the collapse of massive stars is highly uncertain. 
An accurate dynamical mass estimate has been derived only for $\approx{}10$ stellar BHs (see table~2 of \"Ozel et al. 2010 for one of the most updated compilations, but see also Lee, Brown \&{} Wijers 2002; Orosz 2003; Narayan \&{} McClintock 2005). 
Most of the derived BH masses are in the range $6\le{}m_{\rm BH}/{\rm M}_\odot{}\le{}10$, with an apparent absence (in X-ray binaries) of BHs with $m_{\rm BH}<5$ M$_\odot{}$, difficult to explain with observational biases (\"Ozel et al. 2010). In the Milky Way (MW), the most massive BHs in X-ray binaries  do not seem to significantly exceed $m_{\rm BH}\sim{}15$ M$_\odot{}$ (e.g. $m_{\rm BH}=12\pm{}2$ M$_\odot{}$ for GS2023+338, Charles \&{} Coe 2006; $m_{\rm BH}=14\pm{}4$ M$_\odot{}$ for GRS 1915+105, Harlaftis \&{} Greiner 2004), whereas a few BHs in nearby galaxies may have higher masses: M33 X-7 ($m_{\rm BH}=15.65\pm{}1.45$ M$_\odot{}$, Orosz et al. 2007), IC~10 X-1 ($m_{\rm BH}\sim{}23-34$ M$_\odot{}$, Prestwich et al. 2007; Silverman \&{} Filippenko 2008) and NGC 300 X-1 ($m_{\rm BH}>10$  M$_\odot{}$, Carpano et al. 2007; Crowther et al. 2007; $m_{\rm BH}=20\pm{}4$  M$_\odot{}$, under reasonable assumptions for the inclination and for the mass of the companion, Crowther et al. 2010). 

Interestingly, these three relatively massive stellar BHs are hosted in regions with relatively low metallicity. A metallicity $Z=0.22$ Z$_\odot{}$ is estimated for the dwarf irregular galaxy IC 10 from an electron-temperature based calibration of spectra of HII regions ($12+\log{\rm O/H}=8.26\pm{}0.10$, Garnett 1990), assuming Z$_\odot{}=0.019$. The metallicity of M~33 in proximity of  X-7 (i.e. at $\sim{}0.23\,{}R_{25}$, where $R_{25}$ is the Holmberg radius) is $Z\sim{}0.40$ Z$_\odot{}$, and that of NGC~300 in the vicinity of X-1  (i.e. at $\sim{}0.32\,{}R_{25}$) is $Z\sim{}0.28$ Z$_\odot{}$ (derived from the metallicity gradients of M~33 and of NGC~300, respectively, provided by Pilyugin, V\'ilchez \&{} Contini 2004).

From a theoretical perspective, our knowledge of the mass spectrum of stellar BHs is hampered by the uncertainties about two issues: (i) mass losses by stellar winds in massive stars (during and especially after the main sequence, MS); (ii) the hydrodynamics of core-collapse supernova (SN) explosions.

The rate of mass loss by stellar winds during the MS likely increases with the metallicity of the star ($\dot{M}\propto{}Z^{\mu{}}$, where $\mu{}\sim{}0.5-0.9$, depending on the model, e.g. Kudritzki, Pauldrach \&{} Puls 1987; Leitherer, Robert \&{} Drissen 1992; Maeder 1992; Kudritzki \&{} Puls 2000; Vink, de Koter \&{} Lamers 2001; Kudritzki 2002; Belkus, Van Bever \&{} Vanbeveren 2007; Pauldrach, Vanbeveren \&{} Hoffmann 2012).
The behaviour of post-MS massive stars, such as luminous blue variable stars (LBVs) and Wolf-Rayet stars (WRs), is much more uncertain (e.g. Vink \&{} de Koter 2005).

 According to models of stellar evolution and SN explosion (Fryer 1999; Fryer \&{} Kalogera 2001; Woosley, Heger \&{} Weaver 2002; Heger et al. 2003a), a star with a final mass\footnote{We name `final mass', $m_{\rm fin}$, of a star the mass bound to the star immediately before the collapse.} $m_{\rm fin}\gtrsim{}40$ M$_\odot{}$ can collapse quietly to a BH, after a weak (if any) SN explosion (failed SN scenario). In the following, we use the terms `failed SN' and `direct collapse' as synonymous, to indicate the silent collapse of a star to a BH after no or weak SN.
The actual value of the minimum final mass for a star to directly collapse into a BH is quite uncertain: our adopted fiducial value of 40 M$_\odot{}$ is a conservative assumption (Fryer 1999), and searches for SN progenitors provide some evidences for this theoretical scenario (see Smartt 2009 for a recent review).
  Since stellar winds are suppressed at low metallicity, metal-poor massive stars are more likely to have $m_{\rm fin}\gtrsim{}40$ M$_\odot{}$ than their metal-rich analogues. 

In the case of a failed SN, it is reasonable to expect that the mass of the remnant is comparable to the final mass of the progenitor star (or at least more than half of it, see e.g. Heger et al. 2003b). Therefore, BHs may form, via this channel, with mass higher than in the case of SN explosion. According to Belczynski et al. (2010; hereafter B10), the mass of a BH formed via direct collapse may be as high as $\sim{}80$ M$_\odot{}$. The models by B10 consider only the evolution of single, non-rotating stars. Rotation and binarity can affect the final mass of the remnant (e.g. Vanbeveren 2009; Maeder \&{} Meynet 2010, 2012).

 In the following, we will refer to massive stellar BHs (MSBHs) to indicate BHs with mass $25-80$ M$_\odot{}$ formed via direct collapse. The existence of MSBHs in the nearby Universe may be crucial for our understanding of X-ray sources. The scenario of X-ray binaries powered by MSBHs was recently proposed to explain a large fraction of the ultra-luminous X-ray sources (ULXs, i.e. X-ray sources with luminosity, assumed isotropic, higher than $10^{39}$ erg s$^{-1}$), without requiring excessive super-Eddington factors or more exotic mechanisms (e.g. Mapelli, Colpi \&{} Zampieri 2009; Zampieri \&{} Roberts 2009; Mapelli et al. 2010, 2011a).

 In this paper, we present new $N-$body simulations of dense intermediate-mass (a few $\times{}10^3$ M$_\odot{}$) young ($\le{}100$ Myr) star clusters (SCs), including an accurate treatment of dynamics and updated recipes for metal-dependent stellar evolution, stellar winds and failed SNe. Our aim is to study the formation and dynamical evolution of stellar BHs and MSBHs in young SCs with different metallicity. In the current paper (which is the first of a series), we will focus on the effects 
of the dynamics of stellar BHs and  MSBHs 
on the population of X-ray sources. In the next papers of the series we will consider also other effects of BH dynamics (e.g. the consequences for the population of gravitational wave sources).

We simulate SCs, as most stars ($\sim{}80$ per cent) are expected to form in SCs (e.g. Lada \&{} Lada 2003). We restrict our analysis to intermediate-mass (a few $\times{}10^3$ M$_\odot{}$) young SCs. These form with a higher frequency than larger clusters in the nearby Universe (as the mass function of star clusters is ${\rm d}N/{\rm d}m\propto{}m^{-\alpha{}}$, with $\alpha{}\sim{}2$, Lada \&{} Lada 2003), and are often sites of an intense X-ray activity: many bright high-mass X-ray binaries (HMXBs) and ULXs are associated with OB associations and with young intermediate-mass SCs (e.g. Goad et al. 2002;  Zezas et al. 2002; Liu, Bregman \&{} Seitzer 2004; Soria et al. 2005; Ramsey et al. 2006; Terashima, Inoue \&{} Wilson 2006; Abolmasov et al. 2007; Berghea 2009; Swartz, Tennant \&{} Soria 2009; Tao et al. 2011; Gris\'e et al. 2011, 2012). Tens of intermediate-mass young SCs have been discovered in the MW in the last few years (e.g. Bica et al. 2003; Mercer et al. 2005; Borissova et al. 2011; Richards et al. 2012)\footnote{We stress that our results cannot be easily generalized to more massive SCs and in particular to globular clusters, as the masses and the relevant timescales are too different (in globular clusters the half-mass relaxation time is orders of magnitude longer than the timescale for the evolution of massive stars, while in intermediate-mass  SCs these two timescales are comparable). The study of such larger systems requires dedicated simulations.}.


The paper is organized as follows. In Section 2,  we briefly review the previous work on this topic, to summarize the state-of-art and to highlight the differences with the present analysis. In Section 3, we describe the method adopted for the simulations. In Section 4, we present the results, focusing on the mass spectrum of BHs and on the effects of dynamics on accreting BHs in X-ray binaries. In Section 5, we summarize the most relevant results and discuss future challenges for $N-$body simulations of SCs.


\section{Short review of previous work: the importance of combining dynamics and stellar evolution}

Most population synthesis codes study the formation of BHs from single stars and from stars in primordial binaries (i.e. stars that are in the same binary since their formation). This method has been widely used to investigate the population of X-ray binaries (e.g. Portegies Zwart, Verbunt \&{} Ergma 1997; Hurley, Tout \&{} Pols 2002; Podsiadlowski, Rappaport \&{} Pfahl 2002; Podsiadlowski, Rappaport \&{} Han 2003; Belczynski et al. 2004a; Belczynski, Sadowski \&{} Rasio 2004b; Rappaport, Podsiadlowski \&{} Pfahl 2005; Dray 2006; Madhusudhan et al. 2006, 2008; Belczynski et al. 2008; Linden et al. 2010, hereafter L10).
Some  studies include also recipes for the effects of metallicity on stellar evolution (e.g. Hurley et al. 2002; Belczynski et al. 2004a; Dray 2006; Belczynski et al. 2008; L10). 

In particular,  L10 adopt recipes for stellar winds and failed SNe that are very similar to B10, and find that MSBHs are unlikely to power bright X-ray binaries. This occurs because primordial binaries merge if they are sufficiently tight to enter a common envelope (CE) phase before the SN of the primary and if they are still in Roche lobe overflow (RLO) at the end of the CE phase. If they survive the CE phase, they may be tight enough to later enter a RLO phase when the secondary evolves, but typically their BHs are small,  as a consequence of the mass lost during CE. In both cases, the binary does not evolve into a RLO HMXB with a MSBH. 

On the other hand, L10 and the large majority of studies that take into account metallicity do not include the effects of dynamics on the evolution of primordial binaries and on the formation of new binaries. This is a strong limitation, as most stars ($\sim{}80$ per cent) likely form in SCs (e.g. Lada \&{} Lada 2003), and most SCs are collisional environments, that is sites where close encounters between stars and binaries (three-body encounters) are extremely frequent and have important consequences (see, e.g. Bonnell \&{} Kroupa 1998, for a review dedicated to young SCs). Binaries are important for close encounters, because they have a larger cross-section than single stars and because they have an energy reservoir (their internal energy), which can be exchanged with single stars (e.g. Heggie 1975; Heggie \&{} Hut 1993; Davies 1995; Colpi, Mapelli \&{} Possenti 2003). Close encounters with single stars statistically lead to the increase (decrease) of the binding energy of a hard (soft) binary, defined as a binary with binding energy higher (lower) than the average kinetic energy of a star in the SC (Heggie 1975). Close encounters can even unbind binaries (ionization). Dynamical exchanges are also possible, that is interactions where one of the former members of the binary is replaced by the single star (for a description of the possible outcomes of a binary-single star interaction, see, e.g. Sigurdsson \&{} Phinney 1993). Finally, recoil velocities due to three-body encounters
can cause the ejection of the star and/or of the binary from the parent cluster (see Section~4.1 of Sigurdsson \&{} Phinney 1993 for a general definition of recoil velocity in three-body encounters). Recent $N-$body simulations of dense young SCs with MSBH binaries\footnote{In the following, we call MSBH binary (BH binary) a binary hosting at least one MSBH (BH).}, but without stellar-evolution recipes, showed that close encounters substantially affect the evolution of these binaries, inducing hardening, exchanges and even ejections from the parent cluster (Mapelli et al. 2011b). 

Blecha et al. (2006, hereafter B06) is one of the few studies of massive BHs where a stellar evolution code (although not accounting for different metallicities) is combined with recipes for dynamics (in part semi-analytical prescriptions and in part a $N-$body integrator for three-body encounters).  B06 study the evolution of an intermediate-mass BH (IMBH) born via runaway collapse (Portegies Zwart \&{} McMillan 2002) at the centre of a 5$\times{}10^4$ M$_\odot$ dense young SC. 
 The IMBH mass in the simulations of B06 is in the range $50-500$ M$_\odot{}$, i.e. partially overlapping with our definition of MSBHs. B06 find that the dynamical effects are very important for the evolution of the IMBH, as captures of stars and three-body encounters allow it to have a companion star for most of the SC lifetime.

The main differences between this paper and B06 are the following. Firstly, our treatment of dynamics is fully $N-$body, rather than based on a hybrid code. Secondly, in our simulations MSBHs form self-consistently, as a consequence of metallicity-dependent stellar evolution and dynamics, whereas 
 B06 generate one IMBH per cluster, following recipes from runaway-merger simulations.
Thus, `our' MSBHs are generally smaller than those by B06, and do not form necessarily in the core. Thirdly, we included a metallicity-dependent treatment of stellar evolution and wind-mass losses. At last, the total mass of each of the SCs we simulate is a factor of $\gtrsim{}10$ smaller than in B06 simulations. 

 Finally, Monte Carlo codes are suitable for the study of the largest collisional systems ($\sim{}10^5-10^7$ objects), such as globular clusters (e.g. H\'enon 1971a, 1971b; Stod\'olkiewicz 1982, 1986; Giersz 1998; Joshi et al. 2000, 2001; Fregeau et al. 2003; Fregeau \&{} Rasio 2007). The number of objects in these systems makes prohibitive to run wide grids of $N-$body models so far. Some Monte Carlo codes also include accurate recipes for metal-dependent stellar and binary evolution (e.g. Chatterjee et al. 2010; Downing et al. 2010, 2011; Pattabiraman et al. 2012). Monte Carlo codes were recently  used to study the population of X-ray binaries in globular clusters (e.g. Ivanova et al. 2006, 2008), the evolution of possible sources of gravitational waves (e.g. Downing et al. 2010, 2011), and the formation of IMBHs by runaway collapse (e.g. G\"urkan, Freitag \&{} Rasio 2004). These studies do not include recipes for the formation of MSBHs in the $25-80$ M$_\odot$ range.

\section{Method and simulations}

The simulations were done using the Starlab\footnote{\tt http://www.sns.ias.edu/$\sim{}$starlab/} public software environment (see Portegies Zwart et al. 2001), which allows to integrate the dynamical evolution of a SC, resolving binaries and three-body encounters. Starlab includes the SeBa code for stellar and binary evolution (Portegies Zwart \&{} Verbunt 1996; Nelemans et al. 2001). In its original version, SeBa accounts only for a solar-metallicity environment. We modified SeBa by including various effects of metallicity, as follows.

\subsection{Single star evolution}

We added the metallicity dependence of stellar radius, temperature and luminosity, using the polynomial fitting formulas by Hurley, Pols \&{} Tout (2000). We changed the recipes for mass loss by winds for MS stars, by using the metal-dependent fitting formulas given by Vink et al. (2001; see also B10).

We added a very approximate treatment for LBV stars and for WR mass losses by stellar winds, according to the recipes by B10. In particular, we assume that a post-MS star becomes a LBV when its luminosity $L$ and radius $R$ satisfy the requirement that $L/{\rm L}_\odot{}>6 \times 10^5$ and $10^{-5}\,{}(R/{\rm R}_\odot{})\,{}(L/{\rm L}_\odot{})^{0.5} >1.0$ (Humphreys \& Davidson 1994). The mass-loss rate by stellar winds for a LBV is then calculated as $\dot{M} = f_{\rm LBV} \times 10^{-4}$ M$_\odot{}$ yr$^{-1}$, where $f_{\rm LBV}=1.5$ (chosen by B10 because it allows to reproduce the most massive known stellar BHs, see the Introduction and \"Ozel et al. 2010).

Naked helium giants coming from stars with zero age MS (ZAMS) mass $m_{\rm ZAMS}>25$ M$_\odot{}$ (e.g., van der Hucht 1991 and references therein) are labelled as WR stars in the new version of the code and undergo a mass-loss rate by stellar winds defined by $\dot{M} = 10^{-13} (L/{\rm L}_\odot{})^{1.5}\,{}({Z/{\rm Z}_\odot})^{\beta{}}$ M$_\odot{}$ yr$^{-1}$, where $\beta{}=0.86$. This formula was first used by B10, and is  a combination of the Hamann \& Koesterke (1998) wind rate estimate (taking into account WR wind clumping) and Vink \& de Koter (2005) wind $Z$-dependence for WRs.

\begin{figure}
\center{{
\epsfig{figure=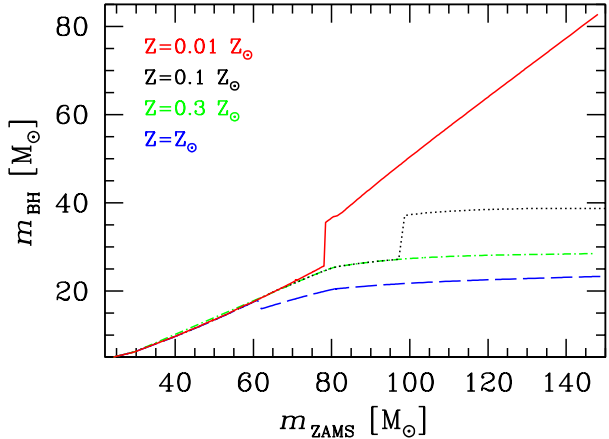,height=6.5cm} 
}}
\caption{\label{fig:fig1}
Mass of the BHs versus ZAMS mass of the progenitor star, as derived by our code when binaries are switched off. Solid line (red on the web): 0.01 Z$_\odot{}$; dotted black line: 0.1 Z$_\odot{}$; dot-dashed line (green on the web): 0.3 Z$_\odot{}$; dashed line (blue on the web): 1 Z$_\odot{}$.}
\end{figure}

 We assume that all stars with final mass $m_{\rm fin}\ge{}40$ M$_\odot{}$ collapse quietly to a BH, without SN explosion (Fryer 1999; Fryer \&{} Kalogera 2001), and that all the BHs born in this way do not receive any natal kicks (see Fryer et al. 2012, for a discussion of this assumption).
If $m_{\rm fin}\ge{}40$ M$_\odot{}$, the mass of the BH is equal to $m_{\rm BH}=m_{\rm CO}+f_{\rm coll}\,{}(m_{\rm He}+m_{\rm H})$, where $m_{\rm CO}$ is the final mass of the Carbon Oxygen (CO) content of the progenitor, while 
$m_{\rm He}$ and $m_{\rm H}$ are the residual mass of Helium (He) and of Hydrogen (H), respectively. $f_{\rm coll}$ is the fraction of He and H mass which collapses to the BH in the failed SN scenario.  The value of $f_{\rm coll}$ is uncertain, and can range between 0 and 1 (depending whether a faint SN or completely no SN occurs). We assume $f_{\rm coll}=2/3$ to match the maximum values of $m_{\rm BH}$ at low $Z$ derived by B10. 

If $m_{\rm fin}<40$ M$_\odot{}$, the SN takes place and we adopt the standard routine of Starlab (Portegies Zwart \&{} Verbunt 1996), including a SN explosion and partial fallback of H, He and CO, depending on the ratio between energy released by the SN and binding energy of each shell.

Fig.~\ref{fig:fig1} shows the mass  of the BH as a function of the ZAMS mass for four different metallicities (from 0.01 to 1 Z$_\odot{}$, where we adopt Z$_\odot{}=0.019$). The values of $m_{\rm BH}$ shown in Fig.~\ref{fig:fig1} have been obtained from our code when binary evolution is switched off. The main feature is the increase of the maximum allowed  $m_{\rm BH}$ for decreasing metallicity. It is interesting to note the abrupt step in  $m_{\rm BH}$ for  $m_{\rm ZAMS}\sim{}80$ and $\sim{}100$ M$_\odot{}$, in the case of $Z=0.01$ Z$_\odot{}$ and $Z=0.1$ Z$_\odot{}$, respectively. The step is produced by the minimum value of $m_{\rm ZAMS}$ for which $m_{\rm fin}\ge{}40$ M$_\odot{}$ and the direct collapse takes place, at a given metallicity.

The main differences between our treatment of the collapse and that by B10 are that (i) B10 do not use the total final mass $m_{\rm fin}$ to discriminate between direct collapse and `standard' SN, but the mass of the CO core, assuming that all stars with final CO mass $m_{\rm CO}\ge{}7.6$ M$_\odot{}$ collapse directly to a BH; (ii) B10 adopt slightly different recipes for fallback (see Portegies Zwart \&{} Verbunt 1996 and Belczynski et al. 2008 for details).

Because of these differences, in our simulations
 fallback is less efficient for ZAMS masses in the $20-40$ M$_\odot$ range.
 Furthermore, the maximum mass of BHs for $Z={\rm Z}_\odot{}$ is higher (by a factor of 1.4) in our simulations with respect to B10 results. Our choice of using  $m_{\rm fin}$  to discriminate between SN and direct collapse produces the abrupt step visible in Fig.~\ref{fig:fig1} for low $Z$, which is much less pronounced in B10.  Our  results are consistent with those of B10 in  light of the uncertainty in the BH mass spectrum. In general, our stellar evolution recipes are less sophisticated than those adopted by B10, but are the most accurate that can be presently implemented within the framework of a complete $N-$body calculation, maintaining an acceptable computational time.

 One of the simplifying approximations in our model (as well as in B10) is that stars do not rotate. Rotation may have important effects both on the mass loss of the progenitor star, inducing rotational mixing (e.g., Maeder \&{} Meynet 2010, 2012, and references therein), and on the formation of the BH, producing asymmetries in the collapse (e.g. Fryer \&{} Heger 2000; Akiyama et al. 2003; Fryer \&{} Warren  2004; Ardeljan, Bisnovatyi-Kogan \&{} Moiseenko 2005; Wheeler, Akiyama\&{}  Williams 2005; Thompson, Quataert \&{} Burrows 2005; Woosley \&{} Bloom 2006). On the other hand, how the effects of rotation affect the mass spectrum of BHs is  debated and highly uncertain (e.g. Dessart, O'Connor, \&{} Ott 2012; Maeder \&{} Meynet 2012). Including the effects of rotation in our model requires a significant revision of the adopted recipes for stellar evolution and wind mass losses, which goes beyond the aims of the current paper. We will investigate the effects of rotation in a forthcoming paper.  

\subsection{Binary evolution}

For the evolution of binaries we maintain the recipes already present in the original version of SeBa (Portegies Zwart \&{} Verbunt 1996).

In particular, SeBa distinguishes among the evolution of detached binaries (where both  stars are smaller than their Roche lobes), that of semi-detached binaries (one of the components fills its Roche lobe) and that of contact binaries (both stars fill their Roche lobe). In the case of detached binaries, mass loss and accretion can occur via stellar winds (according to the formulation by Livio \&{} Warner 1984). In the case of semi-detached binaries, either stable or unstable RLO takes place, according to the formalism described in appendix~C of Portegies Zwart \&{} Verbunt (1996). 
 If the accretor is a BH, the maximum mass-accretion rate of the BH is constrained by the requirement that the luminosity does not exceed the Eddington limit.

A further critical assumption about binary evolution is the CE efficiency (we define as CE efficiency the product of the two degenerate parameters $\alpha{}_{\rm CE}$ and $\lambda{}$, see equation 3 of Podsiadlowski et al. 2003 for a standard definition). For the runs presented in this paper, we adopt $\alpha{}_{\rm CE}\,{}\lambda{}=0.5$, which is a rather standard value and tends to favour the formation of BH binaries (e.g. Podsiadlowski et al. 2003). Test runs with different values of $\alpha{}_{\rm CE}\,{}\lambda{}$ show that  the choice of this parameter does not significantly affect the results for values $\alpha{}_{\rm CE}\,{}\lambda{}\gtrsim{}0.1$, in agreement with Podsiadlowski et al. (2003) and with L10.

\begin{table}
\begin{center}
\caption{Most relevant initial conditions.} \leavevmode
\begin{tabular}[!h]{ll}
\hline
Parameter & Values \\
\hline
$W_0$ & 5 \\
$N_\ast{}$ & 5500 \\
$r_{\rm c}$ [pc] & 0.4\\
$c$ & 1.03\\
IMF & Kroupa (2001)\\
$m_{\rm min}$ [M$_\odot{}$] & 0.1\\
$m_{\rm max}$ [M$_\odot{}$] & 150\\
$f_{\rm PB}$ & 0.1 \\
$Z\,{}[{\rm Z}_\odot{}]$ & 0.01, 0.1, 1.0\\
\noalign{\vspace{0.1cm}}
\hline
\end{tabular}
\begin{flushleft}
\footnotesize{$W_0$: central adimensional potential in the King (1966) model; $N_{\ast}$: number of stars per cluster;  $r_{\rm c}$: initial core radius; $c\equiv{}\log{}_{10}{(r_{\rm t}/r_{\rm c})}$: concentration ($r_{\rm t}$ is the initial tidal radius); $m_{\rm min}$ and $m_{\rm max}$: minimum and maximum simulated stellar mass, respectively; $f_{\rm PB}$: fraction of primordial binaries,  defined as the number of primordial binaries in each SC divided by the number of `centres of mass' (CMs) in the SC. In each simulated SC, there are initially 5000 CMs, among which 500 are designated as `binaries' and 4500 are `single stars' (see Downing et al. 2010 for a description of this formalism). Thus, 1000 stars per SC are initially in binaries.}
\end{flushleft}
\end{center}
\end{table}

\subsection{Initial conditions and simulation grid}
In this paper, we focus on moderately dense SCs, adopting a spherical King profile with central adimensional potential $W_0=5$ (King 1966). Each simulated SC is initially composed of $N_\ast{}=5500$ stars, corresponding to a total mass $M_{\rm TOT}\sim{}3000-4000$ M$_\odot{}$ per SC. The resulting core density, at the beginning of the simulation, is $\rho{}_{\rm c}\sim{}2\times{}10^3$ M$_\odot{}$ pc$^{-3}$. The main parameters adopted for the initial conditions are reported in Table~1. 

For the runs presented in this paper, we fix the primordial binary fraction to $f_{\rm PB}=0.1$. The single stars and the primary stars of each binary are generated according to a Kroupa initial mass function (IMF, Kroupa 2001), with minimum and maximum mass equal to 0.1 and 150 M$_\odot{}$, respectively\footnote{Recent studies show that the IMF might be top-heavy in dense low-metallicity regions (Marks et al. 2012). Thus, our choice of a Kroupa IMF for all the considered metallicities is quite conservative, as it reduces the differences in the number of BHs among different metallicities.}.
The masses of the secondaries ($m_2$) are generated according to a uniform distribution between $0.1\,{}m_1$  and $m_1$ (where $m_1$ is the mass of the primary).  The initial semi-major axis $a$ of a binary is chosen from a distribution $f(a)\propto{}1/a$ (Sigurdsson \&{} Phinney 1995; Portegies Zwart \&{} Verbunt 1996), consistent with the  observations of binary stars in the Solar neighbourhood (e.g. Kraicheva et al. 1978; Duquennoy \&{} Mayor 1991). We generate $a$ between R$_\odot{}$  and $10^5\,{}$R$_\odot{}$, but discarding systems where the distance between the two stars at the pericentre is smaller than the sum of their radii (Portegies Zwart, McMillan \&{} Makino 2007). The maximum value of $a$ was chosen arbitrarily, but motivated by the need to include also a significant fraction of soft binaries\footnote{In the simulated SCs, binaries are soft if $G\,{}m_1\,{}m_2/(2\,{}a)<10^{44}$ erg, where $G$ is the gravitational constant. For $m_1=m_2=1$ M$_\odot{}$ this corresponds to $a\sim{}10^4$R$_\odot$}. 
The initial eccentricity $e$ of a binary is  chosen from a thermal distribution $f(e)=2\,{}e$, in the $0-1$ range (Heggie 1975).

The simulated SCs have half-mass relaxation time $t_{\rm h}\sim{}10\,{}{\rm Myr}\,{}(r_{\rm h}/0.8\,{}{\rm pc})^{3/2}\,(M_{\rm TOT}/3500\,{}{\rm M_\odot})^{1/2}$, where $r_{\rm h}$ is the initial half-mass radius of the SC (in our simulations $r_{\rm h}\sim{}0.8-0.9$ pc). Thus, the core collapse time (Portegies Zwart \&{} McMillan 2002) is $t_{\rm cc}\approx{}2\,{}{\rm Myr}\,{}(t_{\rm h}/10\,{}{\rm Myr})$. We integrate the evolution of these SCs for the first 100 Myr, therefore studying a phase of the life of the cluster in which dynamical interactions are particularly intense.

We make three sets of runs corresponding to three different metallicities: 0.01 Z$_\odot{}$, 0.1 Z$_\odot{}$ and 1 Z$_\odot{}$. For each of these metallicities we simulate 100 different clusters, for a total of 300 SCs.

 The properties of the simulated SCs (total mass, number of stars, core density, core and half-mass radius) are consistent with the properties of observed young intermediate-mass SCs (see e.g. the recent review by Portegies, McMillan \&{} Gieles 2010; see also Hillenbrand \&{} Hartmann 1998; Dias et al. 2002; Portegies Zwart 2004; Pfalzner 2009; Kuhn et al. 2012).

Finally, our simulations do not include recipes for the tidal field of the host galaxy. Accounting for the tidal field may increase the fraction of mass lost, and even transform the SCs into unbound associations (e.g. Gieles \&{} Portegies Zwart 2011). The effect of tidal fields will be added and discussed in forthcoming papers.

 The simulations were run on the graphics processing unit (GPU) cluster IBM-PLX at CINECA. The processors available on PLX are six-cores Intel Westmere 2.40 GHz (two per node), while the GPUs are NVIDIA Tesla M2070 and M2070Q (two per node). Each single job ran over one processor and two GPUs and required 50 CPU hours on average.
Starlab runs on GPUs through the SAPPORO library (Gaburov, Harfst \&{} Portegies Zwart 2009).

\section{Results and discussion}

\subsection{Effects of binary evolution of the progenitor star on BH mass}
\begin{figure}
\center{{
\epsfig{figure=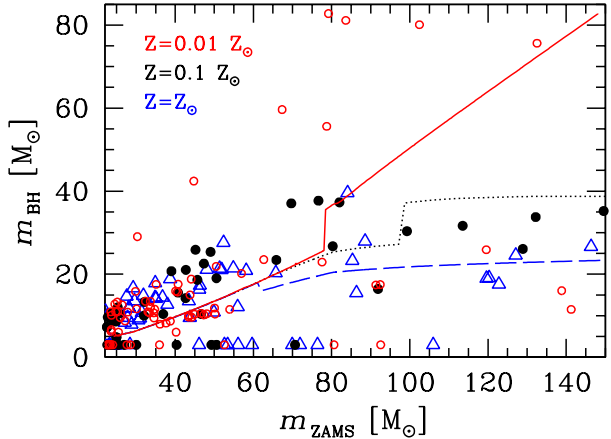,height=6.5cm} 
}}
\caption{\label{fig:fig2}
Mass of the BH versus ZAMS mass of the progenitor star, when binary evolution and dynamics are switched on. Points show the BHs for which $m_{\rm BH}$ differs by more than 5 per cent from the BH mass calculated when the effects of binaries are switched off. Open circles (red on the web): 0.01 Z$_\odot{}$; filled black circles: 0.1 Z$_\odot{}$; open triangles (blue on the web): 1 Z$_\odot{}$. The lines show, for comparison, the behaviour of $m_{\rm BH}$ versus $m_{\rm ZAMS}$ for a population of single stars, and are the same as in Fig.~\ref{fig:fig1}. In particular, solid line (red on the web): 0.01 Z$_\odot{}$; dotted black line: 0.1 Z$_\odot{}$; dashed line (blue on the web): 1 Z$_\odot{}$. }
\end{figure}

 Fig.~\ref{fig:fig2} shows the effects of binary evolution of the progenitor star on the final mass of the BH. In particular, 
the points  in Fig.~\ref{fig:fig2} show  all the simulated BHs (originated from stars in binaries) whose mass differs by more than $5$ per cent from the BH mass calculated if the progenitor was a single star evolving in isolation. In the following, we will simply refer to them as BHs with  $\Delta{}>0.05$. All the BHs  with $\Delta{}>0.05$  form from progenitors that were in a primordial binary and  underwent a mass transfer (MT) phase before collapsing to a BH. The effects of MT before the formation of the first BH can be dramatic, especially when the two stars undergo a CE phase before the first SN. Depending on the binding energy of the envelope, a CE phase can end either with the ejection of the envelope or with the merger of the two stars (see Portegies Zwart \&{} Verbunt 1996 for more details). If the envelope is ejected, the final BH mass will likely be smaller than expected for the evolution of an unperturbed progenitor; whereas if the two stars merge, the final BH mass will be higher than expected for a single star. For simplicity, the code assumes that, if two stars merge during a CE phase, no mass is lost during the merger and that the merger remnant evolves as a MS star (the merger product is a blue straggler star, if its mass is higher than the turn-off mass).

For example, a CE phase followed by envelope ejection is responsible for the two BHs formed by stars with $m_{\rm ZAMS}\sim{}140$ M$_\odot{}$ and $Z=0.01$ Z$_\odot{}$, which have $m_{\rm BH}<20$ M$_\odot{}$, i.e. a factor of $\gtrsim{}4$ less than their analogues born from single stars or wider binaries.

 Interestingly, for $Z=Z_\odot{}$ five BHs form with mass $m_{\rm BH}\ge{}25$ M$_\odot{}$ (one of them with mass close to 40 M$_\odot{}$). These five BHs all form from the merger of a primordial binary before the SN explosion of the primary star\footnote{The formation of MSBHs from the merger of two massive stars at $Z=Z_\odot{}$ depends strongly on the treatment of the star (which is a blue straggler star) formed from the merger, and in particular on the mass lost by this star before the collapse. This issue is delicate and deserves further study.} (see e.g. Soria 2006). The merger is a consequence of the MT phase triggered by the evolution of the primary. This can be a viable path to form MSBHs even at solar metallicity (see the recent paper by Soria et al. 2012 for the case of an ULX powered by a BH with estimated mass $m_{\rm BH}\approx{}40-100$ M$_\odot{}$, in a $\sim{}{\rm Z}_\odot{}$ environment). Alternatively, MSBHs can form at $Z\sim{}{\rm Z}_\odot{}$ even through the merger of two low-mass BHs (e.g. Belczynski et al. 2004b).

Table~2 allows to understand the statistical importance of binary evolution on the final mass of BHs. We define $f_{(\Delta{}>0.05)}$ as the fraction of BHs with $\Delta{}>0.05$ with respect to the total number of simulated BHs. From Table~2, $f_{(\Delta{}>0.05)}\sim{}0.1$ for all the metallicities. Since the primordial binary fraction is $f_{\rm PB}=0.1$, approximately all BHs that formed in primordial binaries have $\Delta{}>0.05$. This depends on the chosen initial distribution of the semi-major axes and on the adopted recipes for CE, but also on the effects of dynamical encounters, as we will discuss in the next Section. This result indicates that the initial binary fraction is an essential ingredient to shape the mass distribution of BHs.

\begin{figure}
\center{{
\epsfig{figure=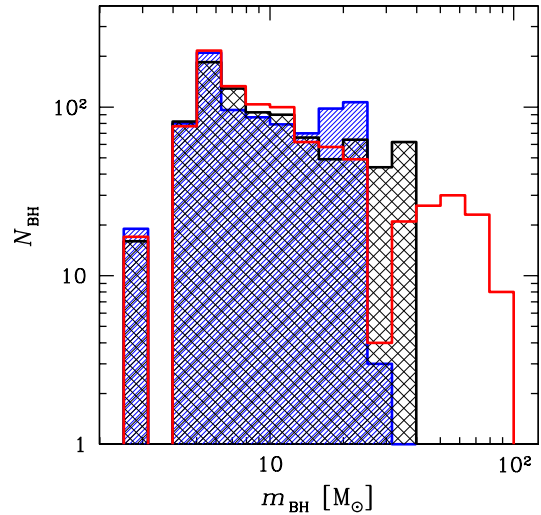,height=7.5cm} 
}}
\caption{\label{fig:fig3}
Mass distribution of BHs in the simulations (including the effect of binaries). Empty histogram (red on the web): 0.01 Z$_\odot{}$; cross-hatched histogram: 0.1 Z$_\odot{}$; hatched histogram (blue on the web): 1 Z$_\odot{}$. BH masses are calculated at the time of formation of the BHs (i.e. do not account for later mergers and/or accretion).}
\end{figure}

The resulting mass distribution of BHs for the three considered metallicities is shown in Fig.~\ref{fig:fig3}. The low-mass tails of the three distributions do not differ significantly, showing a peak at $m_{\rm BH}\sim{}4-6$ M$_\odot{}$.
 This is in fair agreement with the observational limits for MW BHs discussed by \"Ozel et al. (2010), especially considering the uncertainties on fallback models. 

The most significant difference among different metallicities appears at the high-mass tail: the most massive BHs at $Z=1$ Z$_\odot{}$ have a cut-off at $m_{\rm BH}\le{}30$ M$_\odot{}$, whereas a significant fraction of BHs form with $m_{\rm BH}\sim{}40$ M$_\odot{}$ at $Z=0.1$ Z$_\odot{}$ and with $m_{\rm BH}\sim{}40-80$ M$_\odot{}$ at $Z=0.01$ Z$_\odot{}$.


\begin{table*}
\begin{center}
\caption{Statistics of the simulated BHs, when binaries are switched on.} \leavevmode
\begin{tabular}[!h]{llllllllllllll}
\hline
$Z$ [Z$_{\odot}$]
& $N_{\rm BH, cl}$
& $f_{\rm bin}$
& $f_{\rm sin}$
& $N_{\rm exch}$
& $t_{\rm life}$ [Myr]
& $f_{\rm W}$
& $f_{\rm W,\,{}exch}$
& $f_{\rm RL}$
& $f_{\rm RL,\,{}exch}$
& $N_{\rm RL,\,{}int}$
& $f_{(\Delta{}>0.05)}$
& $f_{\rm MSBH}$
& $f_{\rm RL,\,{}MSBH}$
\\
\hline
0.01 & 9.28   & 0.28 & 0.20 & 1.2 & 37.5 & 0.069 & 0.036 & 0.054 & 0.017 & 2.7 & 0.11 & 0.13  &  0.24 \\
0.1  & 8.80   & 0.33 & 0.25 & 1.3 & 31.5 & 0.064 & 0.039 & 0.043 & 0.018 & 2.5 & 0.09 & 0.13  & 0.16 \\
1    & 8.50   & 0.31 & 0.25 & 1.4 & 30.0 & 0.042 & 0.020 & 0.036 & 0.019 & 1.7  & 0.11 & 0.006 & 0.0 \\
\noalign{\vspace{0.1cm}}
\hline
\end{tabular}
\begin{flushleft}
\footnotesize{$N_{\rm BH, cl}$: average number of BHs per cluster; $f_{\rm bin}$: fraction of BHs that are members of a binary at least once in the simulations. This and all the other fractions reported in this Table (except for $f_{\rm RL,\,{}MSBH}$) are calculated with respect to the total number of simulated BHs; $f_{\rm sin}$: fraction of BHs that form from single stars and become members of a binary at least once in the simulations; $N_{\rm exch}$: average number of exchanges per binary that hosts at least one BH (hereafter BH binary); $t_{\rm life}$: average BH binary lifetime; $f_{\rm W}$: fraction of BHs that undergo wind accretion at least once in the simulated time interval; $f_{\rm {W},\,{}exch}$: fraction of BHs that undergo wind accretion with an exchanged companion;  $f_{\rm RL}$: fraction of BHs that undergo RLO at least once in the simulated time interval; $f_{\rm RL,\,{}exch}$: fraction of BHs that undergo RLO with an exchanged companion;  $N_{\rm RL,\,{}int}$:  average number of strong interactions per BH binary (considering only binaries that will undergo RLO); $f_{(\Delta{}>0.05)}$: fraction of BHs whose mass is affected by $>5$ per cent by binary evolution and dynamics;  $f_{\rm MSBH}$: fraction of MSBHs (i.e. BHs with $25\le{}m_{\rm BH}/{\rm M}_\odot{}\le{}80$); $f_{\rm RL,\,{}MSBH}$: fraction of MSBHs that undergo RLO (RLO MSBHs) with respect to the total number of BHs undergoing RLO (RLO BHs).}
\end{flushleft}
\end{center}
\end{table*}

\begin{table*}
\begin{center}
\caption{Statistics of the simulated MSBHs, when binaries are switched on.} \leavevmode
\begin{tabular}[!h]{lllllllllll}
\hline
$Z$ [Z$_{\odot}$]
& $N_{\rm MSBH, cl}$
& $f^{\rm MSBH}_{\rm bin}$
& $f^{\rm MSBH}_{\rm sin}$
& $N^{\rm MSBH}_{\rm exch}$
& $t^{\rm MSBH}_{\rm life}$ [Myr]
& $f^{\rm MSBH}_{\rm W}$
& $f^{\rm MSBH}_{\rm W,\,{}exch}$
& $f^{\rm MSBH}_{\rm RL}$
& $f^{\rm MSBH}_{\rm RL,\,{}exch}$
& $N^{\rm MSBH}_{\rm RL,\,{}int}$
\\
\hline
0.01 & 1.18   & 0.86 & 0.75 & 1.5 & 52.2 & 0.203 & 0.195 & 0.102 & 0.102 & 4.4 \\
0.1  & 1.12   & 0.87 & 0.75 & 1.8 & 44.6 & 0.134 & 0.125 & 0.054 & 0.054 & 4.5 \\
1    & 0.05  & 0.80 & 0.00 & 2.5 & 41.1 & 0.000 & 0.000 & 0.000 & 0.000 & 0.0 \\
\noalign{\vspace{0.1cm}}
\hline
\end{tabular}
\begin{flushleft}
\footnotesize{$N_{\rm MSBH, cl}$: average number of MSBHs per cluster; $f^{\rm MSBH}_{\rm bin}$: fraction of MSBHs that are members of a binary at least once in the simulations. This and all the fractions reported in this Table are calculated with respect to the total number of simulated MSBHs; $f^{\rm MSBH}_{\rm sin}$: fraction of MSBHs that form from single stars and become members of a binary at least once in the simulations; $N^{\rm MSBH}_{\rm exch}$: average number of exchanges per binary that hosts at least one MSBH (hereafter MSBH binary); $t^{\rm MSBH}_{\rm life}$: average MSBH binary lifetime; $f^{\rm MSBH}_{\rm W}$: fraction of MSBHs that undergo wind accretion at least once in the simulated time interval; $f^{\rm MSBH}_{\rm {W},\,{}exch}$: fraction of MSBHs that undergo wind accretion with an exchanged companion;  $f^{\rm MSBH}_{\rm RL}$: fraction of MSBHs that undergo RLO at least once in the simulated time interval; $f^{\rm MSBH}_{\rm RL,\,{}exch}$: fraction of MSBHs that undergo RLO with an exchanged companion;  $N^{\rm MSBH}_{\rm RL,\,{}int}$:  average number of strong interactions per MSBH binary (considering only binaries that will undergo RLO).}
\end{flushleft}
\end{center}
\end{table*}

\subsection{Statistical properties of single and binary BHs}

Table~2 provides a striking evidence of the importance of dynamics on the evolution of BHs in binaries, although the statistics is still quite low. The average number of  BHs per cluster ($\sim{}9$) is consistent with the expectations for a Kroupa IMF and does not appreciably depend on the metallicity.

The fraction of BHs that became members of a binary at least once in the simulated time interval is $f_{\rm bin}\sim{}0.3$, regardless of the metallicity. This fraction is higher (by a factor of $\approx{}$2) than the fraction we expect if all the BH binaries come from primordial binaries. Furthermore, the fraction of BHs that formed from single stars and then became members of a binary as a consequence of a dynamical exchange is very high ($f_{\rm sin}\sim{}0.2$). In particular, $f_{\rm sin}$ is very similar to $f_{\rm bin}$, indicating that most of the BHs in binaries formed from single stars. In addition, the average number of exchanges per BH binary ($N_{\rm exch}$) is more than one during the simulated time interval. Therefore, dynamical exchanges dominate the life of binary BHs in the simulated young SCs.

The average lifetime of BH binaries is $t_{\rm life}\sim{}30-40$ Myr in Table~2. In calculating  $t_{\rm life}$, we assume that a binary survives even if one of its members exchanges, and dies only when it is completely ionized or when the simulation is stopped.
 It is worth noting that BH$-$BH and  BH$-$neutron star (NS)  binaries (i.e. binaries where both the primary and the secondary member are compact objects) live longer than other BH binaries (on average), because they are not perturbed by the stellar evolution of the secondary and are sufficiently massive to avoid ionization.  In particular, the average lifetime of BH$-$BH binaries is 46, 37 and 32 Myr at $Z=0.01$, 0.1 and 1 Z$_\odot{}$, respectively. The average lifetime of BH$-$NS  binaries is similar: 49, 38 and 65 Myr at $Z=0.01$, 0.1 and 1 Z$_\odot{}$, respectively\footnote{The long average lifetime ($65$ Myr) of BH$-$NS binaries at $Z=1$ Z$_\odot{}$ is explained by statistical fluctuations: only four BH$-$NS binaries form at $Z=1$ Z$_\odot{}$ in our simulations.}. We notice that the lifetime of BH$-$BH and BH$-$NS binaries is significantly longer at very low metallicity ($Z=0.01$ Z$_\odot{}$). At this metallicity, the most massive MSBHs ($m_{\rm BH}>40$ M$_\odot$) tend to produce very massive, hard and thus long-lived binaries. We will focus on binaries composed of two compact objects in a dedicated paper.

The fraction of BHs that undergo wind accretion ($f_{\rm W}$) indicates that there is about one wind-accreting system every two young clusters in 100 Myr. 
Interestingly, about half of these systems are consequences of dynamical exchanges (see $f_{\rm W,\,{}exch}$ in Table~2). This is true for all the considered metallicities (for $Z=1$ Z$_\odot{}$ the fraction of wind-accreting systems formed by exchanges is slightly lower than for lower metallicities, but this might be a fluctuation due to low statistics). 


The fraction of BHs in RLO ($f_{\rm RL}$) is  lower than $f_{\rm W}$.
A large fraction of  RLO systems are originated from a dynamical exchange (in Table~2, $f_{\rm RL,\,{}exch}\sim{}0.3-0.5\,{}f_{\rm RL}$). 
Most of the donor stars in RLO systems are post-MS stars ($\sim{}80$ per cent post-MS stars, versus $\sim{}20$ per cent MS stars). A non-negligible fraction of such post-MS stars in  RLO systems are LBV and WR stars ($\sim30$ per cent of the total donor stars in RLO systems). These results are in agreement with B06. We note that a non-negligible fraction of MS companions are blue straggler stars, i.e. stars rejuvenated by stellar mergers or by MT (e.g. Mapelli et al. 2004, 2006): they behave as MS stars, although they have mass higher than the turn-off mass.


In Table~2, $N_{\rm RL,\,{}int}$ is defined as the average number of strong resonant interactions per binary (i.e. interactions that lead to the formation of an unstable triple system and that change significantly the orbital period of the binary), calculated only for those binaries that will undergo RLO. These interactions are mostly three-body encounters and, in a few cases, four-body encounters (i.e. binary-binary encounters). The number of strong interactions is quite high ($N_{\rm RL,\,{}int}\sim{}2-3$), confirming the importance of three-body encounters.

 Finally, the last two columns of Table~2 show the number of MSBHs normalized to the total number of BHs ($f_{\rm MSBH}$), and the number of  RLO MSBHs normalized to the total number of RLO BHs ($f_{\rm RL,\,{}MSBH}$), respectively. $f_{\rm MSBH}$ and $f_{\rm RL,\,{}MSBH}$ are similar, suggesting that the incidence of RLO systems among MSBHs is comparable to the incidence of RLO systems among low-mass BHs. 

 We stress that the main results presented in this Section, and especially the statistics of accreting systems, strongly depend on the assumed fraction of primordial binaries. In fact, primordial binaries have at least two important effects on accreting systems. Firstly, a number of primordial binary systems are expected to be sufficiently tight to start RLO at early times, as a consequence of stellar evolution. If there are no primordial binaries, the first RLO systems will appear later in the evolution of the SC, as an effect of dynamical interactions. 

Secondly, primordial binaries (and especially hard primordial binaries) represent an initial reservoir of binding energy. 
This means that, if a BH becomes member of a hard primordial binary after an exchange, this binary might be sufficiently tight to start RLO immediately after the first exchange. On the contrary, non-primordial binaries that form from the encounter of three single stars are initially quite soft, and even the hardest among them become sufficiently tight to start RLO only after a lot of dynamical encounters (e.g. Hut et al. 1992). These arguments are discussed quantitatively (through a supplementary set of runs) in Appendix~A.

We also note that the results presented in this Section strongly depend on the stellar evolution recipes and on the simplified model of failed SN adopted for our simulations. In Appendix~B, we highlight the differences with respect to a scenario where the formation of MSBHs is strongly suppressed. In the forthcoming investigations, we will consider in more detail different models of stellar evolution.

\subsection{Statistical properties of single and binary MSBHs}
Table~3 reports the same quantities as shown in the first 11 columns of Table~2, but calculated only for the MSBHs (rather than for all the BHs).
Table~3 shows that dynamical interactions are crucial for MSBHs, even more important than for low-mass stellar BHs. 

First, the fraction of MSBHs that became members of a binary at least once in the simulated time interval (with respect to the total number of simulated MSBHs) is $f^{\rm MSBH}_{\rm bin}\sim{}0.8-0.9$, i.e. a factor of $\sim{}3-4$ higher than in the overall BH sample. At low metallicity  ($Z=0.01$ and 0.1 Z$_\odot{}$), most of the MSBHs in binaries formed from single stars, as the fraction of MSBHs that formed from single stars and then became members of a binary as a consequence of a dynamical exchange is $f^{\rm MSBH}_{\rm sin}=0.75$. Instead, for $Z=1$ Z$_\odot{}$, no MSBHs in binaries form from single stars ($f^{\rm MSBH}_{\rm sin}=0$). This is naturally explained by the fact that the only channel to form MSBHs at $Z=1$ Z$_\odot{}$ is the merger between two massive stars in a primordial binary (see Section 4.1).

Furthermore, the average number of exchanges per MSBH binary is $N^{\rm MSBH}_{\rm exch}\sim{}1.5-2.5$, higher than for the overall BH sample ($N_{\rm exch}=1.2-1.4$).
The lifetime of MSBH binaries is slightly longer than that of the entire BH binary sample ($t^{\rm MSBH}_{\rm life}\sim{}40-50$ Myr). This is in fair agreement with B06, finding that a BH with $m_{\rm BH}=50-100$ M$_\odot{}$,  born from runaway collapse, spends about 25 to 60 per cent of the simulated time interval with a companion.

Columns $7-10$ of Table~3 provide a key result to understand the formation of X-ray binaries powered by MSBHs. The comparison between columns $7$ and $8$ indicates that the large majority of wind-accreting MSBH binaries at low $Z$ ($Z=0.01$ and 0.1 Z$_\odot{}$) were originated by a dynamical exchange ($f^{\rm MSBH}_{\rm W,\,{}exch}\gtrsim{}0.9\,{}f^{\rm MSBH}_{\rm W}$). The comparison between columns $9$ and $10$ indicates that all RLO MSBH binaries at low metallicity ($Z=0.01$ and 0.1 Z$_\odot{}$) were originated by a dynamical exchange ($f^{\rm MSBH}_{\rm RL,\,{}exch}=f^{\rm MSBH}_{\rm RL}$). Furthermore, all but one of the RLO MSBH binaries in our simulations host MSBHs that formed from single stars (this is not shown in Table~3, but we checked it from our simulations). The only MSBH (among RLO MSBH binaries) that did not form from a single star was originated from the merger of a primordial binary. Thus, even this MSBH formed as a single object and acquired a stellar companion after dynamical exchange.

The fraction of wind-accreting MSBH binaries (with respect to the total number of simulated MSBHs) is higher than the corresponding fraction for all BH binaries. The fraction of RLO MSBH binaries is similar to the corresponding fraction for all BH binaries, but the low statistics makes fluctuations very important (we find 18 RLO MSBH binaries over 300 simulations). Therefore, to study  RLO MSBH binaries in detail, it will be essential to run a much larger grid of simulations (Mapelli et al., in preparation).

We find no wind-accretion MSBH binaries and no RLO MSBH binaries for $Z=1$ Z$_\odot{}$, but this is likely an effect of the low statistics, as only five MSBHs form in our simulations at  $Z=1$ Z$_\odot{}$.

The results reported in Table~3 are in fair agreement with the findings by L10, based on population synthesis calculations. In particular, L10 find that no or very few RLO systems powered by MSBHs form from primordial binaries. This is consistent with our finding that no RLO MSBH binaries come from primordial binaries. On the other hand, the code used by L10 does not account for dynamical interactions. All RLO MSBH binaries found in our simulations were originated from dynamical interactions, in the sense that the MSBH became a member of the binary as a consequence of a dynamical exchange.

\begin{figure*}
\center{{
\epsfig{figure=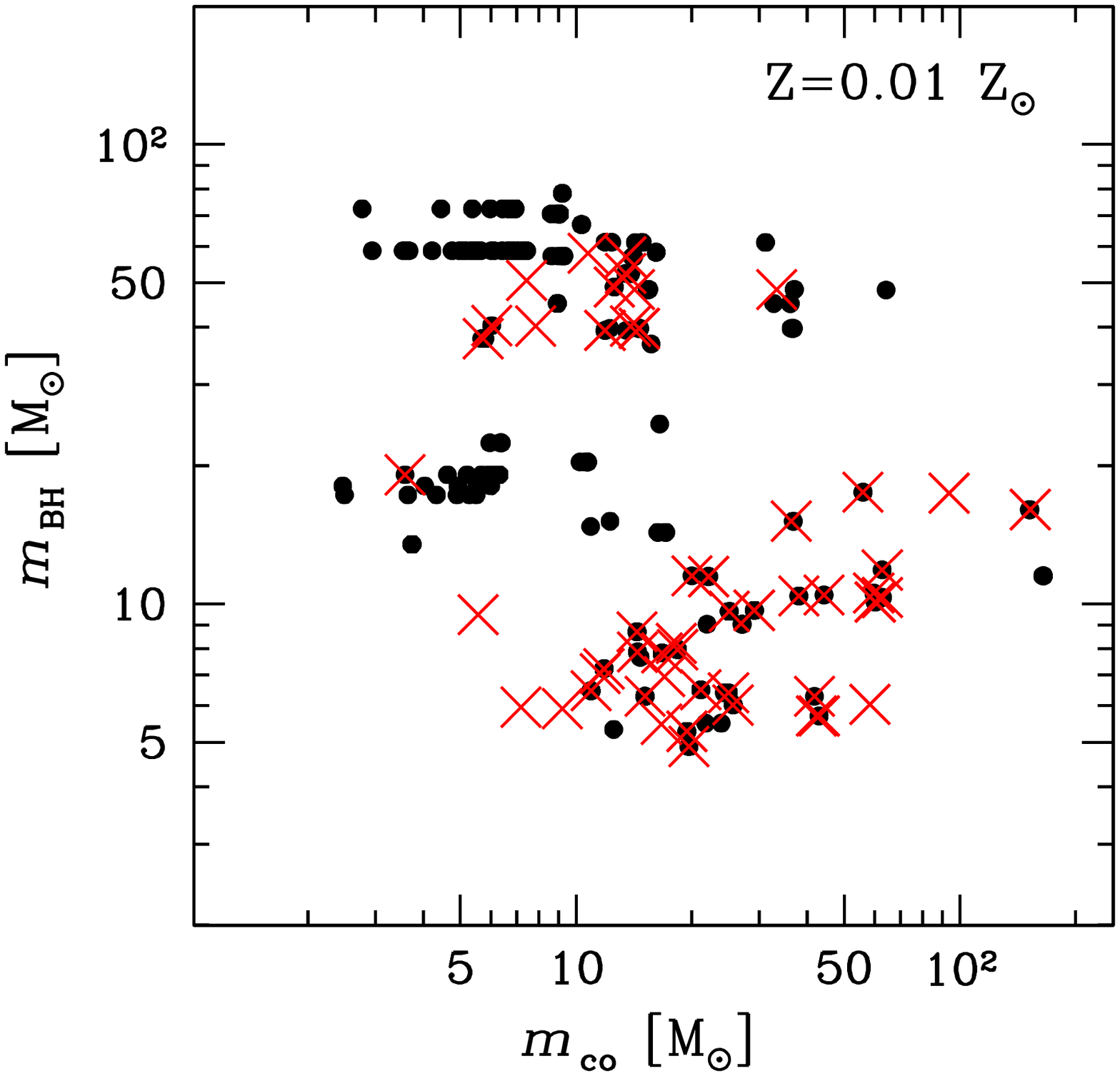,height=5.5cm} 
\epsfig{figure=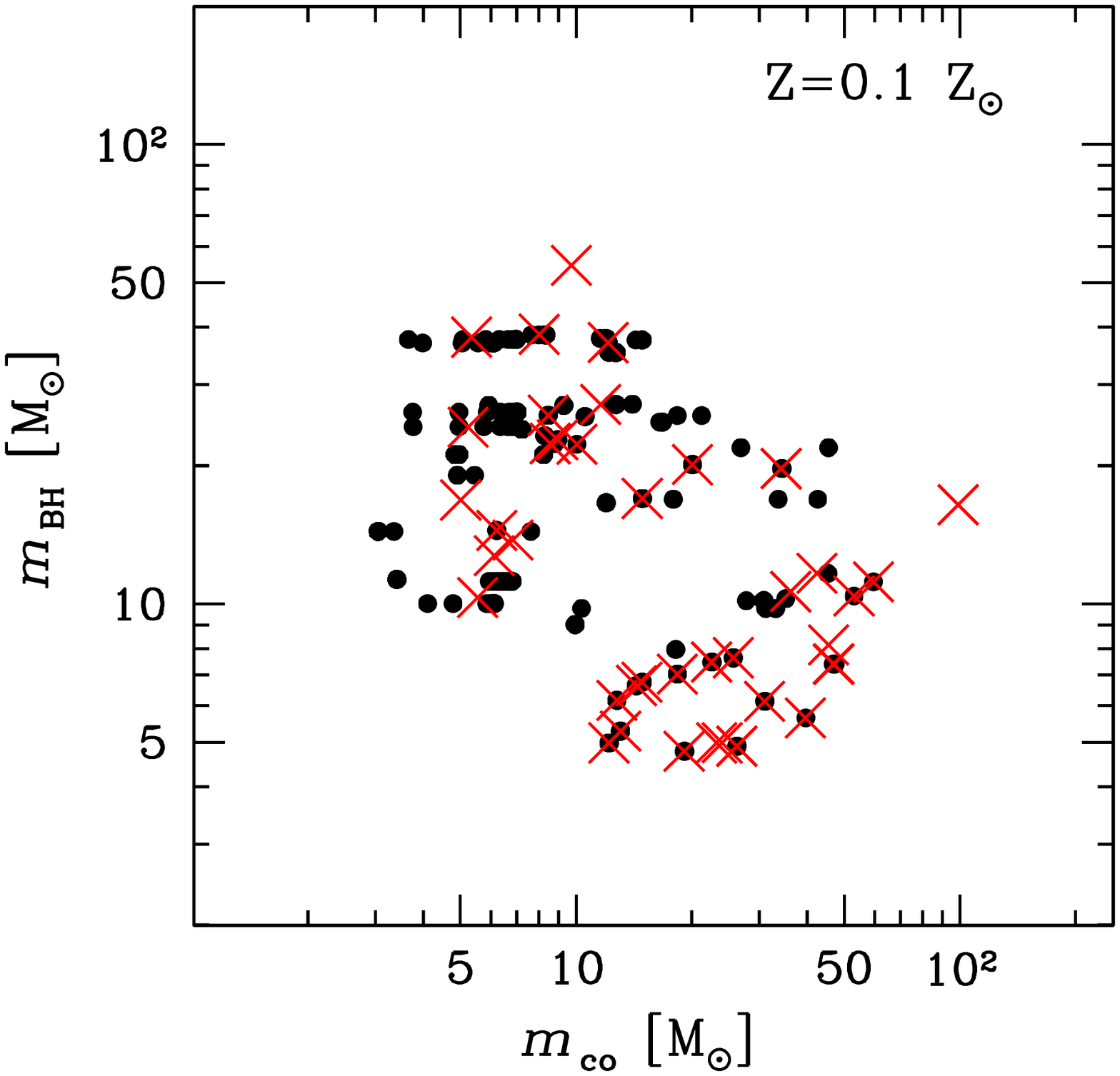,height=5.5cm}
\epsfig{figure=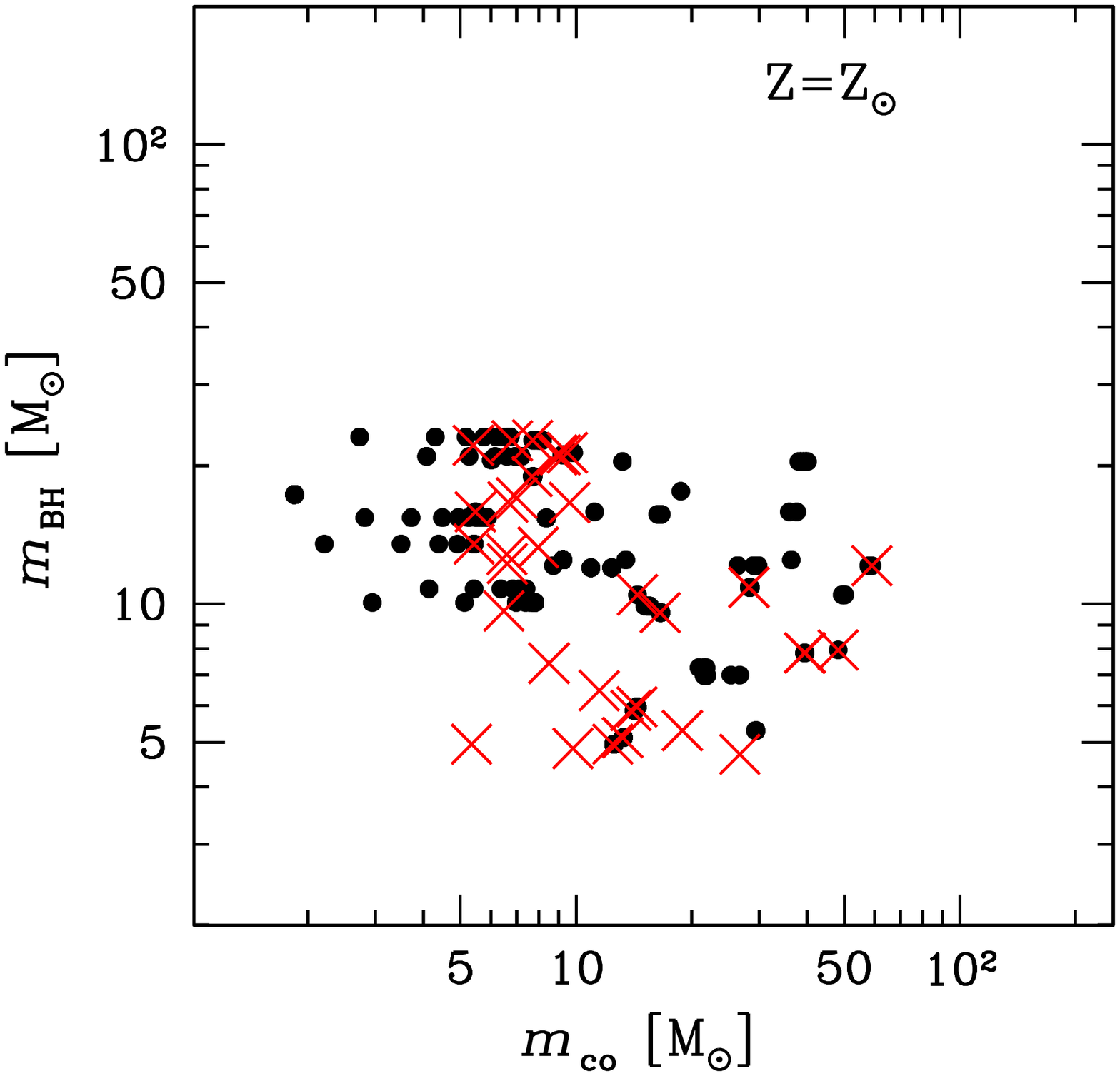,height=5.5cm}  
}}
\caption{\label{fig:fig4}
Mass of the BH versus mass of the companion star. Filled circles: wind-accretion systems; crosses (red on the web): RLO systems (at the first RLO epoch). From left to right:  0.01 Z$_\odot{}$,  0.1 Z$_\odot{}$, 1 Z$_\odot{}$. Each system can be identified by more than one point, when the mass of the secondary changes significantly (because of mass losses or because of dynamical exchange). A cross and a circle almost superimposed indicate that the same system passes from wind-accreting to RLO (or {\it vice versa}).}
\end{figure*}

\begin{figure*}
\center{{
\epsfig{figure=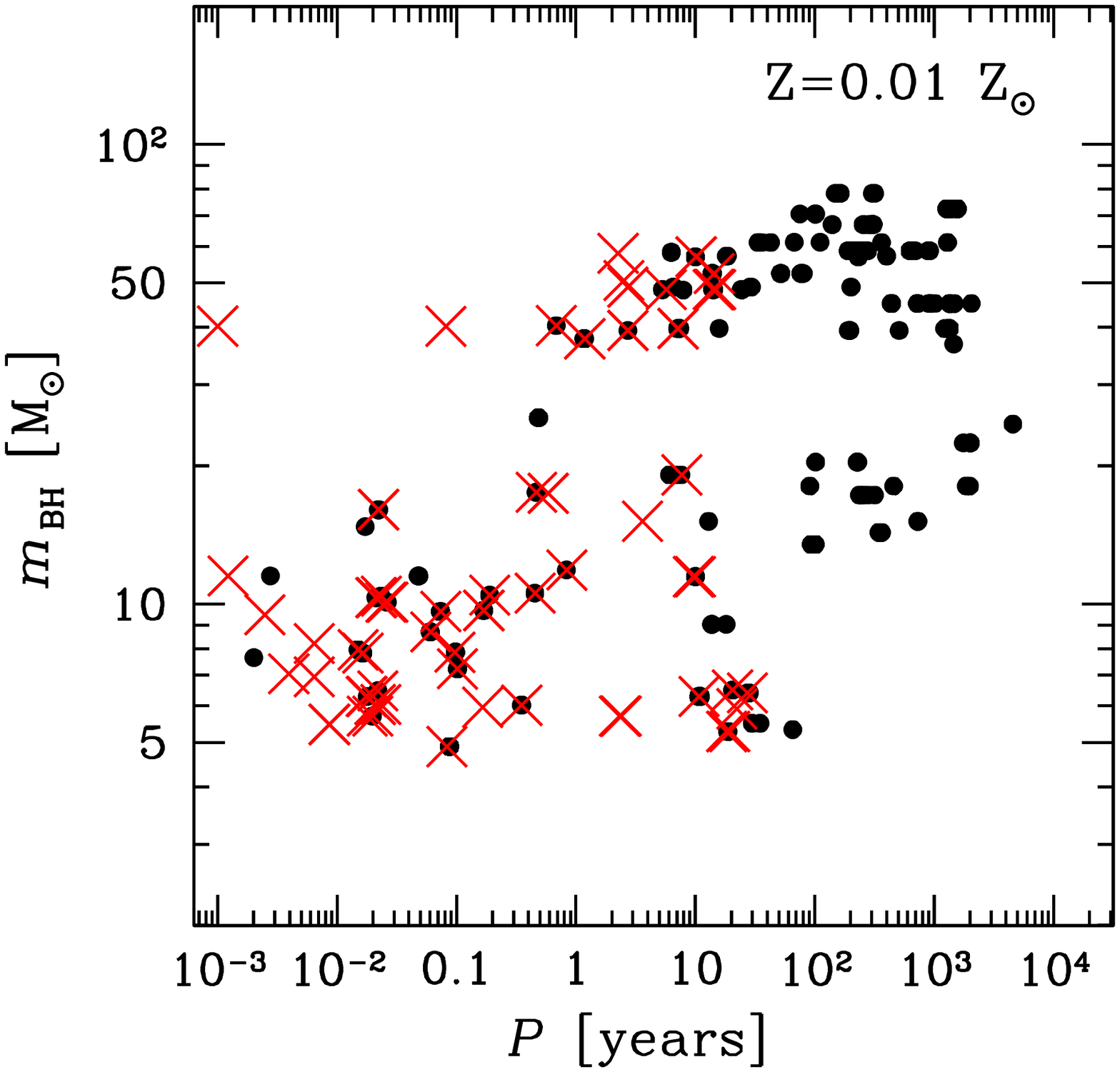,height=5.5cm} 
\epsfig{figure=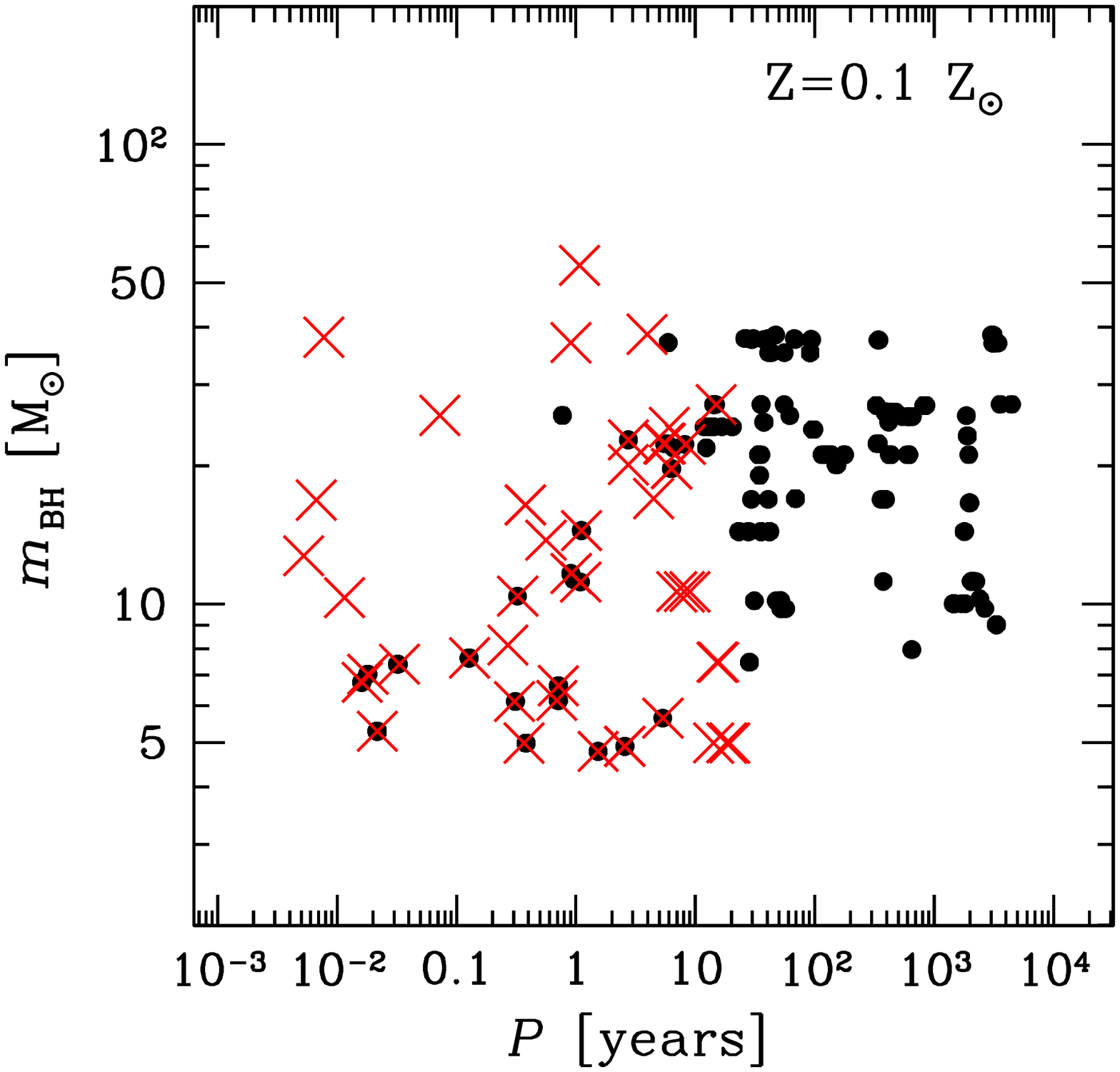,height=5.5cm}
\epsfig{figure=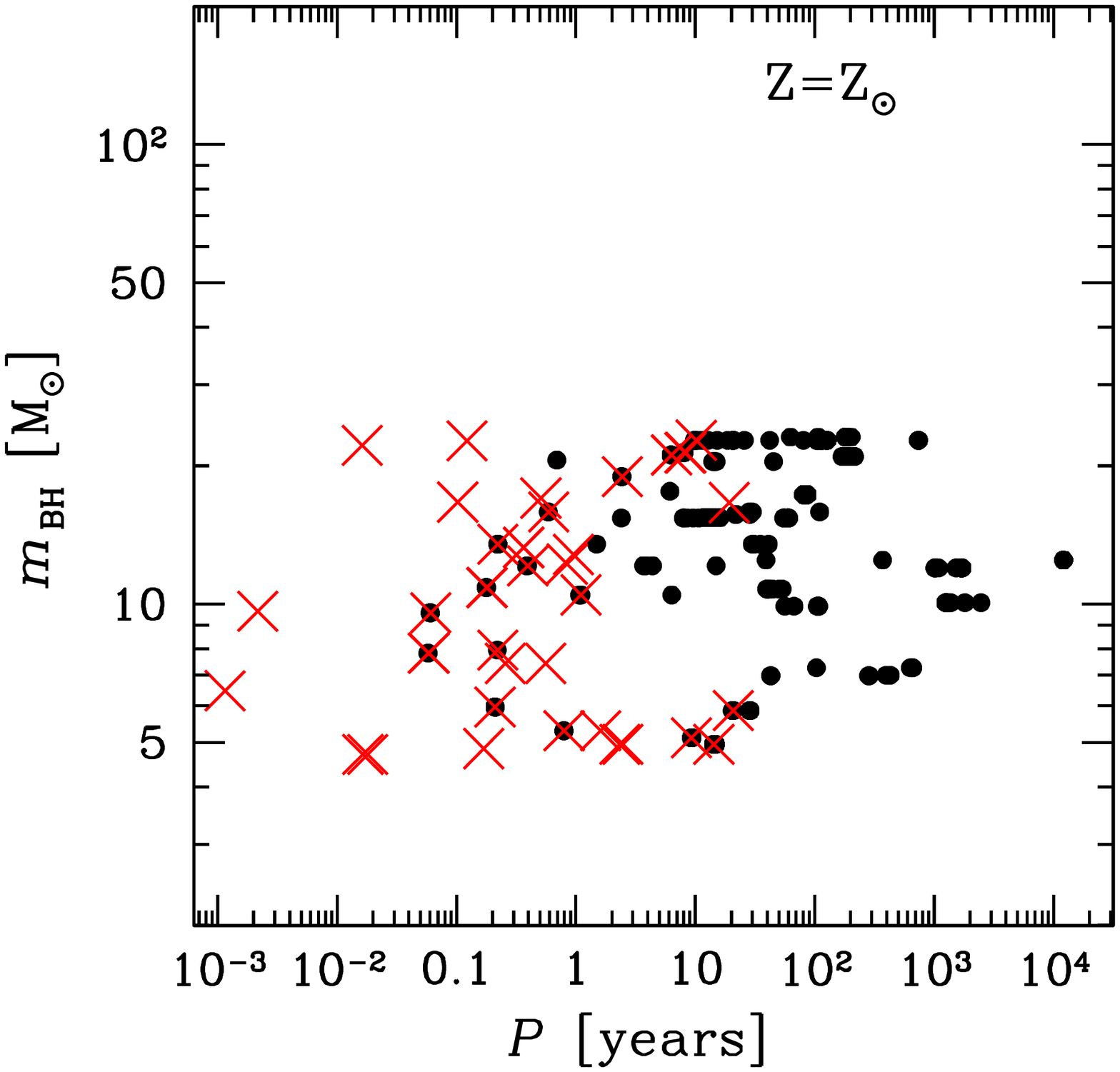,height=5.5cm}  
}}
\caption{\label{fig:fig5}
Mass of the BH versus orbital period. Filled circles: wind-accretion systems; crosses (red on the web): RLO systems (at the first RLO epoch). From left to right:  0.01 Z$_\odot{}$,  0.1 Z$_\odot{}$, 1  Z$_\odot{}$. Each system can be identified by more than one point, when the period evolves significantly, as consequence of accretion, circularization, or dynamical interactions.}
\end{figure*}

\begin{figure*}
\center{{
\epsfig{figure=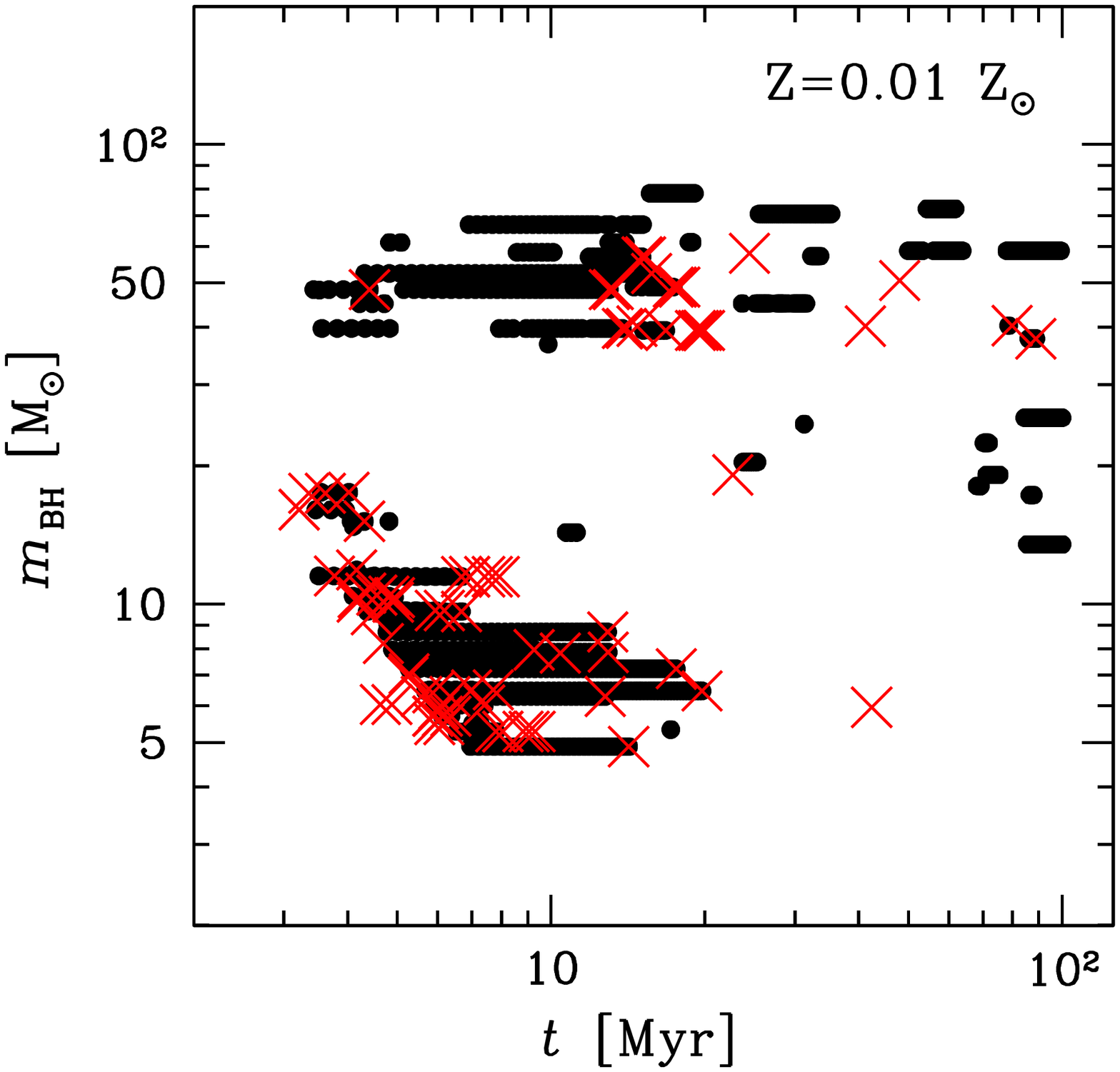,height=5.5cm} 
\epsfig{figure=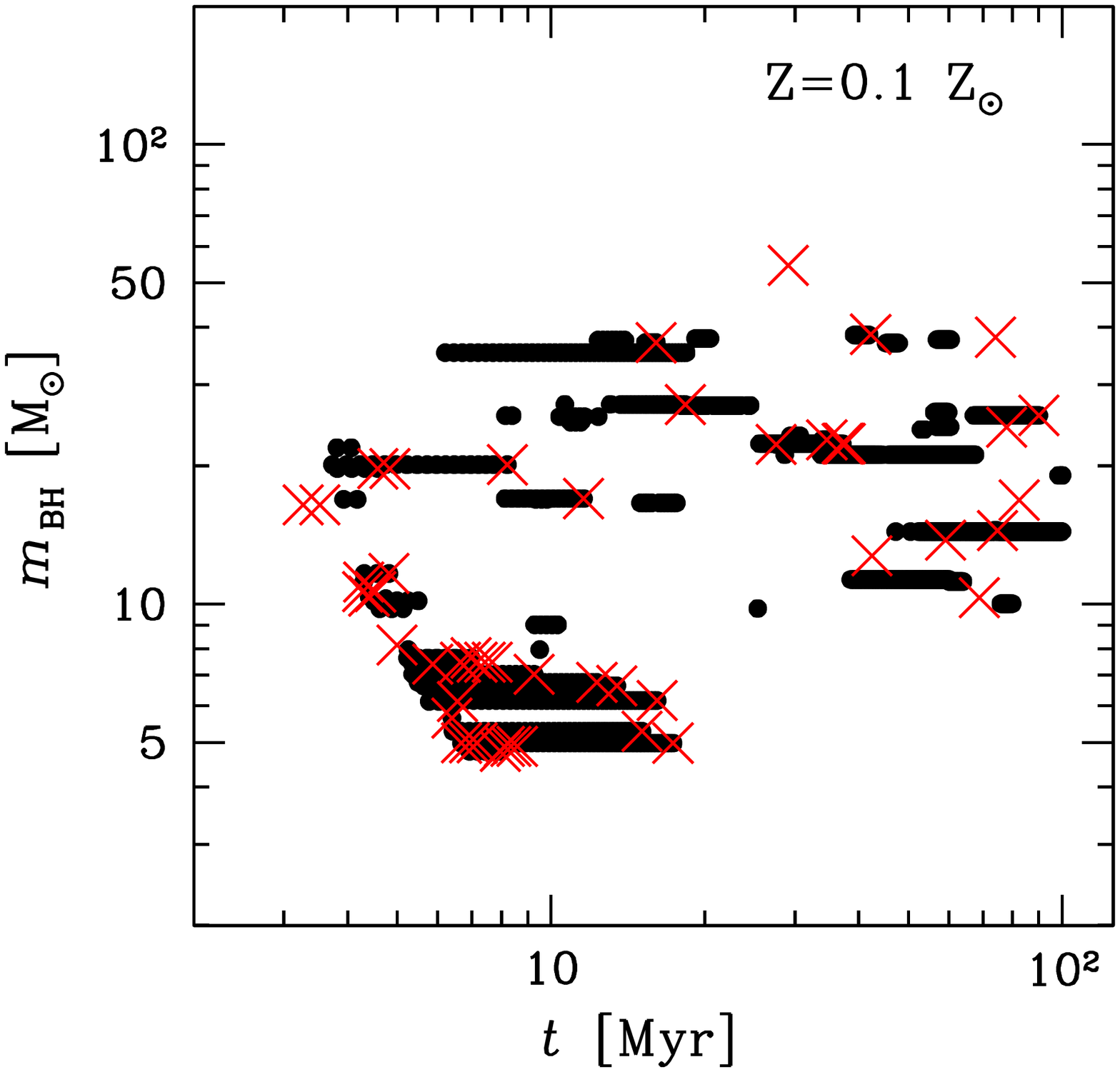,height=5.5cm}
\epsfig{figure=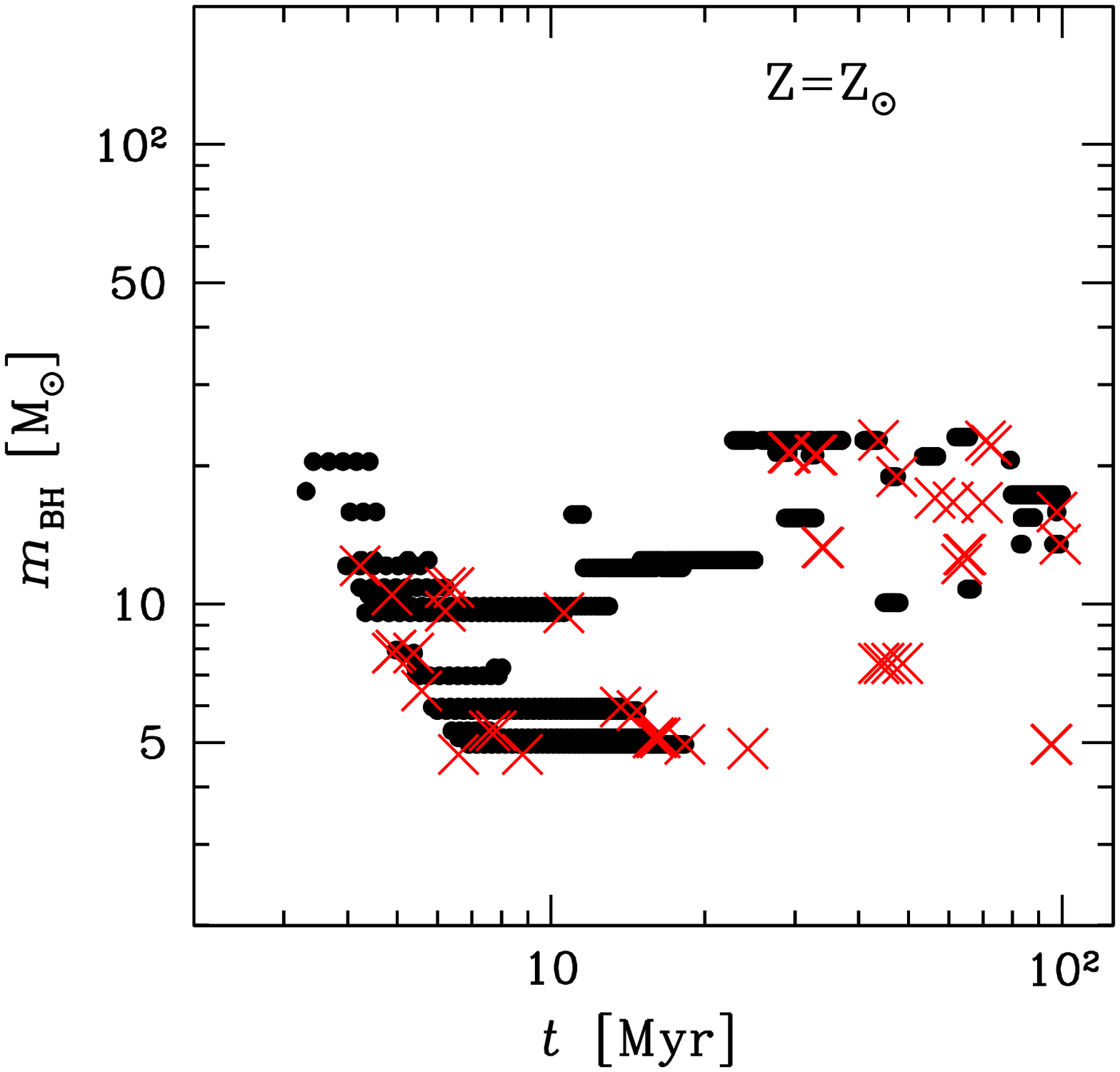,height=5.5cm}  
}}
\caption{\label{fig:fig6}
Mass of the BH versus time elapsed since the beginning of the simulation.
 Filled circles: wind-accretion systems; crosses (red on the web): RLO systems (at the first RLO epoch). From left to right:  0.01 Z$_\odot{}$,  0.1 Z$_\odot{}$, 1 Z$_\odot{}$.  Each circle (cross) in this Figure indicates that the binary was wind accreting (in RLO) during a single snapshot. Therefore, each system can be identified by more than one point, when the MT lasts for more snapshots (the time between two snapshots being $\sim{}0.26$ Myr).}
\end{figure*}

\subsection{Orbital properties of accreting BHs}
Figs.~\ref{fig:fig4}, \ref{fig:fig5} and \ref{fig:fig6} show some important characteristics of MT BHs. In these Figures, the filled circles are wind-accreting systems, while the crosses are RLO systems. 
Fig.~\ref{fig:fig4} shows the mass of the BH versus the mass of the companion star ($m_{\rm co}$). 
At all considered metallicities, $m_{\rm co}$ ranges from relatively low values ($\sim{}2$ M$_\odot{}$, mostly asymptotic giant branch stars) to very high values ($\sim{}60$ M$_\odot{}$ or more) for both wind-accreting systems and RLO systems. 
The masses of accreting BHs span all the possible BH masses for a given metallicity (see Fig.~\ref{fig:fig3}, for comparison), even the higher masses. 
 Not only wind-accreting systems but also RLO systems can host MSBHs ($m_{\rm BH}\sim{}25-60$ M$_\odot{}$). This is an important difference with respect to L10 results, and it is mainly the consequence of dynamical interactions (which were not accounted by L10, see the discussion in Section 4.3).
 We note that for $Z=0.1$ Z$_\odot$ there is even one RLO system powered by a $\sim{}55$ M$_\odot{}$ MSBH. This MSBH is more massive than the maximum mass ($\sim{}40$ M$_\odot{}$) that can be reached at this metallicity through the adopted models of stellar evolution. In fact, such MSBH comes from the merger between a $\sim{}7$ M$_\odot{}$ BH and a  $\sim{}50$ M$_\odot{}$ MS companion.

Fig.~\ref{fig:fig5}  shows the mass of the BH versus the orbital period $P$. 
For all the considered metallicities, wind-accreting systems can form with periods as long as a few $\times{}10^6$ days, in agreement with previous studies (e.g. L10). 

RLO systems have periods spanning from less than one day ($8.8$ hours for one system at $Z=0.01$ Z$_\odot$) up to $\sim{}10$ yr. We stress that the periods of RLO systems shown in Fig.~\ref{fig:fig5} are the values of the period at the first Roche lobe approach, as the code does not trace with accuracy the late stages of RLO and the time interval between snapshots is not sufficiently short to follow the evolution of all the systems. For example, if a system evolves into tidal instability and the companion is a MS star, the code assumes that the system is undergoing merger and removes the binary from calculation (Portegies Zwart \&{} Verbunt 1996). Therefore, the plotted periods of RLO systems must be considered upper limits. We note that the companion star is an evolved star, with a very large radius (of the order of $100$ R$_\odot$), in most of the wide RLO systems ($P\ge{}1$ yr).

Fig.~\ref{fig:fig6} shows the mass of the BH versus the time elapsed since the beginning of the simulation. 
This plot gives information about the duty cycle ($t_{\rm duty}$), defined as the lapse of time for which a binary is wind accreting and/or in RLO, divided by the total elapsed time in the simulation.
Fig.~\ref{fig:fig6} indicates that most systems are in RLO for less than one snapshot (corresponding to $\sim{}0.26$ Myr). This is in fair agreement with the results by B06, which find that RLO systems powered by IMBHs are on average short-lived ($\lesssim{}1$ per cent of the simulated time interval). Wind accretion can last for a longer time (a few Myr). 

Relatively low-mass BHs ($<15$ M$_\odot{}$) tend to start the RLO phase short after their formation. These systems start RLO as a consequence of stellar evolution of the secondary and/or as an effect of the first SN kick (see L10). Instead, more massive BHs start RLO at later times ($10-90$ Myr after the beginning of the simulation). This means that the most massive BHs were single or in relatively wide binaries and can start RLO only as a consequence of the hardening of the binary by three-body encounters and/or of a dynamical exchange. 
Therefore, dynamical interactions are essential to allow MSBHs to power RLO X-ray binaries. 
 This difference is also important to understand which X-ray sources are associated with low-mass BHs and which X-ray sources might be powered by MSBHs. In fact, from Fig.~\ref{fig:fig6} we expect  that the donor stars in X-ray binaries powered by MSBHs are on average older than those in X-ray binaries powered by low-mass BHs. In particular, most MSBHs enter the RLO phase with companions that are $\approx{}10-50$ Myr old. Interestingly, most of the ULXs for which information about the stellar environment is available are associated with $\sim{}10-30$ Myr old stellar populations (see e.g. Soria et al. 2005; Liu et al. 2007; Gris\'e et al. 2008; Swartz et al. 2009; Gris\'e et al. 2011; Voss et al. 2011).

\begin{figure}
\center{{
\epsfig{figure=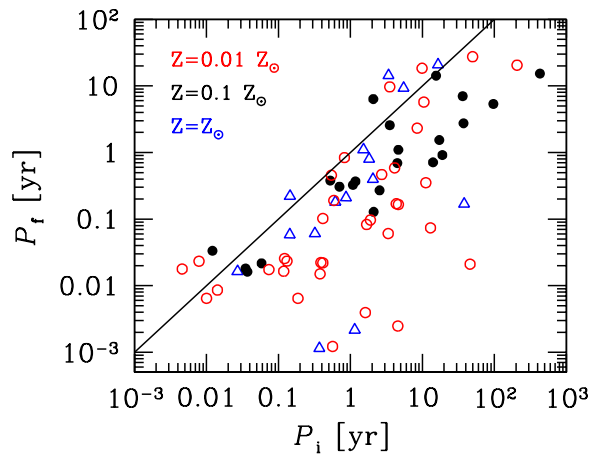,height=6.5cm} 
}}
\caption{\label{fig:fig7}
Period at the first RLO versus the initial period of the binary (only for those binaries that do not undergo exchange). Open circles (red on the web): 0.01 Z$_\odot{}$; filled black circles: 0.1 Z$_\odot{}$; open triangles (blue on the web): 1 Z$_\odot{}$. The solid line marks the points with $P_{\rm f}=P_{\rm i}$.}
\end{figure}

Further hints about the importance of dynamics come from the variation of the orbital period. Fig.~\ref{fig:fig7} shows the final period $P_{\rm f}$ of a RLO binary (defined as the period at the beginning of RLO) versus the initial period $P_{\rm i}$ of the same binary (defined as the period at the beginning of the simulation). For consistency, in Fig.~\ref{fig:fig7} we show only primordial binaries that do not undergo dynamical exchanges before starting RLO. For most systems $P_{\rm f}<P_{\rm i}$, as it was reasonable to expect. The shrinking is due to the joint effect of stellar evolution (e.g. a CE phase before the formation of the first BH forces the semi-major axis to shrink) and of three-body encounters (especially when the period changes by a factor of $\sim{}10$ or more). 

The large majority of systems have $P_{\rm i}>0.01$ yr. Systems with an initial period below this threshold merge before the formation of the first BH. There are only two systems with $P_{\rm i}<0.01$ yr, both for $Z=0.01$ Z$_\odot{}$. These systems undergo at least one strong three-body encounter before the formation of the first BH. The three-body encounters widen the semi-major axis of these binaries, allowing them to avoid merger and to survive till the formation of the first BH\footnote{According to Heggie's law (Heggie 1975), hard binaries tend to harden as a consequence of three-body encounters. This is true in a statistical sense. Single interactions can widen even hard binaries.}.


\subsection{BH ejections}
\begin{table}
\begin{center}
\caption{Statistics of the BH ejections.} \leavevmode
\begin{tabular}[!h]{lllllll}
\hline
$Z$ [Z$_{\odot}$]
& $f_{\rm ej}$
& $f_{\rm ej,\,{}SN}$
& $f_{\rm ej,\,{}MSBH}$
& $f_{\rm ej,\,{}bin}$
& $f_{\rm ej,\,{}RL}$
\\
\hline
0.01 &  0.50 & 0.23  & 0.03  & 0.10 & 0.006 \\ 
0.1  &  0.41 & 0.19  & 0.04  & 0.09 & 0.009 \\ 
1    &  0.46 & 0.24  & 0.001 & 0.08 & 0.007 \\ 
\noalign{\vspace{0.1cm}}
\hline
\end{tabular}
\begin{flushleft}
\footnotesize{$f_{\rm ej}$: fraction of BHs (including MSBHs) that are ejected from the SC (i.e. that have distance from the centre of the cluster $>2\,{}r_{\rm t}$). This and all the other fractions reported in this Table are normalized to the total number of simulated BHs; $f_{\rm ej,\,{}SN}$: fraction of BHs that are ejected by natal kick (i.e. after the SN explosion); $f_{\rm ej,\,{}MSBH}$: fraction of MSBHs that are ejected from the SC;  $f_{\rm ej,\,{}bin}$: fraction of BHs that are ejected from the SC together with their companion star; $f_{\rm ej,\,{}RL}$: fraction of BHs that are ejected from the SC together with their companion star and start RLO after the ejection.}
\end{flushleft}
\end{center}
\end{table}
 The possibility that a BH is ejected from the parent SC is relevant for various astrophysical issues. 
 For example, bright HMXBs and ULXs are often close to young SCs and star forming regions, but displaced by $\sim{}10-1000$ pc with respect to their centre (e.g. Zezas et al. 2002; Kaaret et al. 2004; Berghea 2009; Swartz et al. 2009; Swartz 2010; Rangelov, Prestwich \&{} Chandar 2011; Voss et al. 2011; Poutanen et al. 2012). This fact has been generally interpreted as the indication that bright X-ray sources are powered by runaway binaries, that is, by binaries that were ejected from the parent cluster because of a natal kick (e.g. Sepinsky, Kalogera \&{} Belczynski 2005; Zuo \&{} Li 2010) or because of a close encounter (e.g. Kaaret et al. 2004; Berghea 2009; Mapelli et al. 2011b).

In our simulations, BHs can be ejected both through SN explosion (natal kick) and through three-body encounters. Table~4 shows that $\sim{}40-50$ per cent of simulated BHs are ejected from the SC, almost independently of the metallicity (we classify a BH as ejected when its distance from the centre of mass of the SC is $>2\,{}r_{\rm t}$). About half of the ejections are consequences of the natal kick, while the remaining half is due to three body encounters. SN explosions can also unbind a binary system. We estimate that $\sim{}0.2$  primordial binaries per SC are ionized by a SN explosion leading to the formation of a BH (we do not include NSs in this estimate).

At $Z=0.01$ and 0.1 Z$_\odot{}$, the fraction of ejected MSBHs (with respect to the total number of simulated BHs) is $f_{\rm ej,\,{}MSBH}\sim{}0.03-0.04$.
Since MSBHs are $\sim{}13$ per cent of all the BHs at these metallicities, this means that about $20-30$ per cent of all the simulated MSBHs are ejected as a consequence of three-body encounters (we recall that in our simulations we assume that MSBHs receive no natal kick). This percentage is moderately lower, but still in agreement with Mapelli et al. (2011b), who find that $\sim{}40$ per cent of all MSBHs are ejected as a consequence of three-body encounters. The difference between these two estimates can be explained by the fact that  Mapelli et al. (2011b) generate the MSBHs already in the initial conditions (rather than letting them form later, through stellar evolution), and assume that all MSBHs are members of primordial binaries. Thus, the available time for MSBHs to undergo thee-body encounters is longer in the simulations by Mapelli et al. (2011b).
This assumption increases the number of interactions involving MSBH binaries, especially during the first stage of core collapse (which occurs as early as $2-3$ Myr in our simulated SCs). Furthermore, the simulations in Mapelli et al. (2011b) do not include stellar evolution.

Table~4 also shows that about 10 per cent of all the simulated BHs are ejected together with their companion star, regardless of the metallicity. About one tenth of these ejected binaries enter RLO after leaving the cluster ($f_{\rm ej,\,{}RL}\sim{}0.006-0.009$). Since the total fraction of RLO systems is $f_{\rm RL}\sim{}0.04-0.05$ (see Table~2), this means that about 20 per cent of all RLO systems enter the RLO phase after being ejected from the SC (and in most of the cases the RLO is triggered by a dynamical interaction). Only one of these ejected RLO systems is powered by a MSBH. Thus, most of the simulated RLO systems are inside the parent SC. This result is apparently at odds with observations (e.g. Zezas et al. 2002; Kaaret et al. 2004; Berghea 2009; Swartz et al. 2009; Swartz 2010), which indicate that a significant fraction of bright X-ray sources are offset from the parent cluster by more than 10 pc. On the other hand, we stress that the simulations presented in this paper do not include the tidal field of the host galaxy: as our simulated SCs have a relatively small initial mass, a large fraction of binaries can be stripped from the SC by external tidal forces. 

Finally, in the current paper we have assumed that the MSBHs are born without natal kick. This assumption depends crucially on the model of direct collapse and failed SN. Thus, the fraction of ejected MSBHs reported in this paper must be regarded as a lower limit.
\begin{figure*}
\center{{
\epsfig{figure=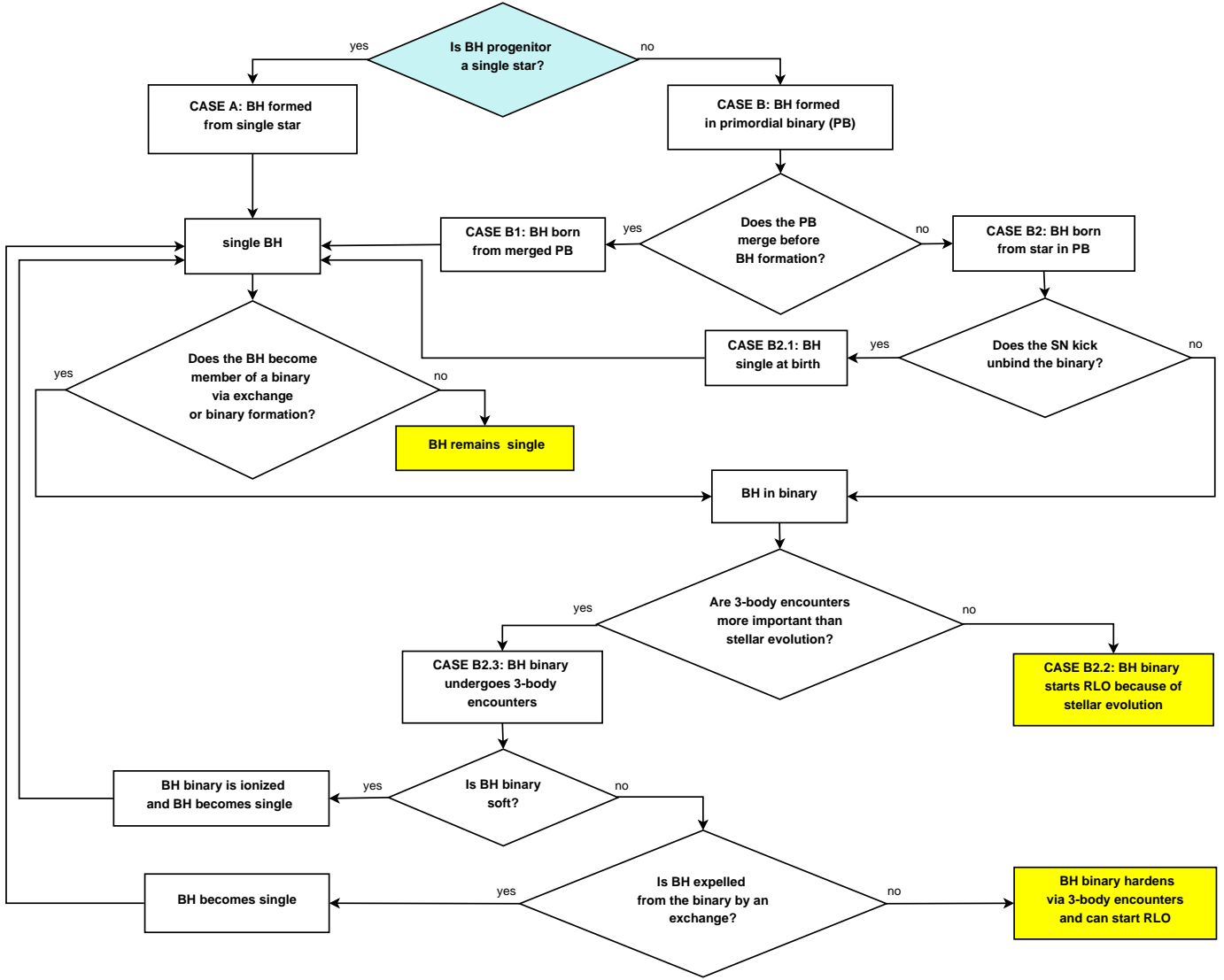,height=15.0cm} 
}}
\caption{\label{fig:fig8}
Flow chart summarizing the BH evolution, in light of the interplay between stellar evolution and dynamics. The starting point and the end points of the flow chart are highlighted.}
\end{figure*}

\subsection{A schematic interpretation of MSBH behaviour}
 In the previous sections, we highlighted the differences between the behaviour of MSBHs and that of low-mass stellar BHs, discussing the results of the $N-$body simulations. In this section, we show how the results of the $N-$body simulations can be intuitively understood in light of the interplay between stellar evolution and some of the basic properties of three-body encounters.
The three most relevant aspects from the physics of three-body encounters are the following.

(i) The probability of a single star (or stellar remnant) with mass $m_3$ to exchange into a binary ($\mathcal{P}_{\rm exch}$) is higher if $m_3>m_1$ or $m_3>m_2$, where $m_1$ and $m_2$ are the masses of the two components of the binary (e.g. Hills \&{} Fullerton 1980; Sigurdsson \&{} Phinney 1993). In particular, $\mathcal{P}_{\rm exch}$ is very close to one, if the binary is hard (so that it cannot be easily ionized) and if $m_3>2\,{}m_1$ or $m_3>2\,{}m_2$ (Hills \&{} Fullerton 1980). After the first $\approx{}10$ Myr of the cluster lifetime (when the turn-off mass goes below $\approx{}15$ M$_\odot{}$), single MSBHs are among the most massive objects in the SC and are likely to replace the lowest-mass members of primordial binaries.

(ii) The cross section  for  three-body interactions  is larger for more massive binaries (see e.g. Sigurdsson \&{} Phinney 1993; Davies, Benz \&{} Hills 1994; Davies 2002; Miller \&{} Hamilton 2002). After the first $\approx{}10$ Myr of the cluster lifetime,  MSBH binaries become significantly more massive than the other binaries in the SC. Thus, their rate of three-body encounters is higher than that of low-mass binaries.

(iii) Hard binaries tend to harden as a consequence of three-body encounters, i.e. to reduce their semi-major axis $a$ (e.g. Heggie 1975; Heggie \&{} Hut 1993; Davies 1995; Quinlan 1996; Merritt 2001). The decrease of $a$ can start RLO into a BH binary. A MSBH binary in our simulated SCs is hard if $a\lesssim{}10^3$ A.U.$\,{}(m_{\rm BH}/25\,{}{\rm M}_\odot)$ (corresponding to a period $P\lesssim{}10^4$ yr$\,{}(m_{\rm BH}/25\,{}{\rm M}_\odot)$). Therefore, most of the MSBH  binaries in our simulations are hard. 

Combining stellar evolution with the above notions from the theory of three-body encounters, we can summarize the evolution of the simulated BHs as shown in Fig.~\ref{fig:fig8}. In the following, we discuss the main differences between MSBHs and low-mass BHs, relatively to the flow chart in Fig.~\ref{fig:fig8}.

Case A: the BH forms from a single star. If the BH is a MSBH, then the exchange probability $\mathcal{P}_{\rm exch}$ is very high already after $\sim{}10$ Myr since the SC formation, and the MSBH becomes soon a binary member via dynamical exchange. The resulting MSBH binary undergoes efficiently three-body encounters and hardens rapidly.

If the BH has relatively low mass, its $\mathcal{P}_{\rm exch}$ is low for most of the SC life: the low-mass BH can remain single for a much longer time, or even for the entire lifetime of the cluster. Furthermore, low-mass BHs are more easily ejected out of the SC. 

Case B: the BH forms from a primordial binary. Depending on the initial semi-major axis of the binary and on the stellar evolution, the binary can merge or avoid merger before the formation of the BH.  If the binary merges before the formation of the BH (Case B1), then the BH forms as a single object and its evolution is the same as in the Case A. 

If the binary does not merge before the formation of the BH (Case B2), then there are at least three possibilities.

Case B2.1: The SN kick of the primary unbinds the binary. In this case, the BH is substantially a single object and behaves as in Case~A. In our simulations, this case can occur only for low-mass BHs, as MSBHs are assumed to form without natal kick.

Case B2.2: The binary remains bound and the stellar evolution of the secondary is faster than three-body encounters. This occurs especially when the binary is very close and/or the mass of the secondary is similar to the mass of the primary.
For binaries where the radius of the secondary star is close to the Roche lobe at the time of formation of the BH, the binary enters a RLO phase because of the evolution of the secondary. As already shown by L10, MSBHs can hardly form in such close binaries, because the first CE phase leads generally to the formation of small BHs.  Therefore, this case is more frequent for low-mass BHs.


Case B2.3: The binary remains bound and three-body encounters are more efficient than the stellar evolution of the secondary. This occurs especially for relatively wide binaries (where stellar evolution is not sufficient to drive RLO), and/or for low-mass secondary stars, which evolve much more slowly than the primary. In this case, the evolution is very different depending on the mass of the BH and depending on whether the binary is hard or soft. In the following, we mention only three of the possible cases (the most relevant for our simulations).

(i) If the binary is soft (unlikely if a MSBH is member of the binary), three-body encounters are expected to ionize it.

(ii) If the BH is a MSBH and the binary is relatively hard, then three-body encounters harden the binary. Exchanges can occur, but are unlikely to remove the MSBH from the binary, as it is more massive than most single stars. The MSBH binary survives and  may undergo MT.

(iii) If the BH has a low mass (lower than the companion mass) and the binary is relatively hard,  then three-body encounters harden the binary, but exchanges can remove the BH from the binary. If it is expelled by an exchange, the BH becomes single again.


We stress that the flow chart in Fig.~\ref{fig:fig8} is schematic and somehow simplistic, as it neglects some further effects that can take place (e.g. the SN explosion of the companion star, the formation of a binary composed of two compact objects, the definitive ejection of the BH from the SC). Despite this,  Fig.~\ref{fig:fig8} 
 allows to understand why exchanges are so important to enhance the formation and evolution of MSBH binaries (see Table~3). In fact, MSBHs evolve predominantly from Case~A (i.e. they form from single stars) to Case B2.3,  where exchanges and three-body encounters dominate the evolution of a hard and massive MSBH binary. 


It clarifies also why most low-mass BHs start RLO immediately after their formation (see Fig.~\ref{fig:fig6}), whereas MSBHs enter RLO much later than their formation ($\gtrsim{}20$ Myr after the beginning of the simulation). In fact, it is unlikely for MSBHs to evolve through Case B2.2 and to start RLO because of the stellar evolution of the secondary star in a primordial binary. MSBHs evolve mainly through Case A and Case B2.3: they can efficiently acquire (new) companions because of dynamical exchanges, and their binaries harden because of three-body encounters. On the other hand, dynamical interactions occur on a longer timescale (a few tens Myr) than massive star evolution.

 Instead, low-mass BHs evolve mainly through Case A, Case B2.2 and Case B2.3. In Case A, low-mass BHs remain likely single objects, as the probability of an exchange into a binary is low.
 Case B2.2 can lead to MT on the timescale of massive star evolution. Case B2.3 can result in the ejection of the low-mass BH from the binary, as a consequence of dynamical exchanges. Therefore, most of the RLO systems powered by low-mass BHs evolve as described in Case B2.2, and only a few of them through Case B2.3.

\subsection{RLO systems and Eddington luminosity}
To describe the evolution of mass accretion and the emission properties of RLO binaries is beyond the scope of this paper. On the other hand, from the simulations described in the previous sections, we can extract some basic hints about the evolution of RLO binaries.

The code we use limits the accretion rate so that the luminosity of the binary cannot exceed the Eddington limit for BHs. We prefer not to change this assumption and not to introduce arbitrary rules for super-Eddington accretion, as the dynamical code is not sufficiently accurate to distinguish between different accretion models. In a forthcoming paper, we will re-simulate the MT systems obtained from this study, without accounting for three-body encounters but with a more accurate recipe for the MT rate (e.g. Patruno \&{} Zampieri 2008).

\begin{figure}
\center{{
\epsfig{figure=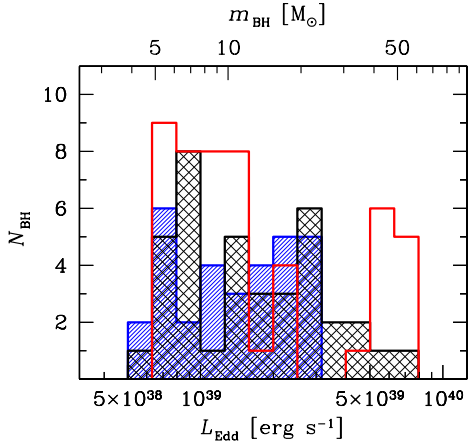,height=6.5cm} 
}}
\caption{\label{fig:fig9}
Distribution of the Eddington luminosity (bottom $x-$axis) and of the BH mass (top $x-$axis) of the simulated RLO systems. Empty histogram (red on the web): 0.01 Z$_\odot{}$; cross-hatched histogram: 0.1 Z$_\odot{}$; hatched histogram (blue on the web): 1 Z$_\odot{}$.}
\end{figure}

Fig.~\ref{fig:fig9} shows the Eddington luminosity ($L_{\rm Edd}$) for the simulated RLO systems. There are important differences between the solar-metallicity simulations and the two sub-solar environments, although the statistics is quite low. 
No RLO systems formed at  $Z={\rm Z}_\odot$ have $L_{\rm Edd}\ge{}2.95\times{}10^{39}$ erg s$^{-1}$ (corresponding to a BH mass $m_{\rm BH}=23$ M$_\odot$). RLO systems formed at $Z=0.1\,{}{\rm Z}_\odot$ and $Z=0.01\,{}{\rm Z}_\odot$ 
reach $L_{\rm Edd}\ge{}7.1\times{}10^{39}$ erg s$^{-1}$ and $L_{\rm Edd}\ge{}7.5\times{}10^{39}$ erg s$^{-1}$, respectively (corresponding to BH mass $m_{\rm BH}=55$ and 58 M$_\odot$, respectively).

\section{Conclusions}
The mass spectrum of BHs born from the collapse of massive stars is very uncertain. Recent theoretical studies (e.g. Mapelli et al. 2009; Zampieri \&{} Roberts 2009; B10) indicate that the mass of the BH depends on the metallicity of the progenitor star and that metal-poor massive stars can produce MSBHs with a mass as high as $\approx{}80$ M$_\odot{}$ (for  $Z=0.01\,{}{\rm Z}_\odot$). Observational results suggest that low-metallicity environments host more massive BHs than the MW (e.g. Prestwich et al. 2007) and  highlight a possible anti-correlation between  the population of bright X-ray sources (especially ULXs) and the metallicity of the host galaxy (e.g. Mapelli et al. 2010, 2011a).

However, it is still unclear whether theoretically predicted MSBHs are efficient in powering X-ray sources. Studies based on population synthesis codes (e.g. L10), including metallicity effects but neglecting dynamical interactions, find that MSBHs can hardly power bright X-ray sources, because of various effects connected with binary evolution.

In this paper, we present preliminary simulations run with a $N-$body plus stellar and binary evolution code, based on the public version of Starlab. These simulations include metallicity-dependent stellar evolution, binary evolution and an accurate treatment of three-body encounters. We simulate dense young SCs with three different metallicities: $Z=0.01\,{}{\rm Z}_\odot$, $Z=0.1\,{}{\rm Z}_\odot$ and $Z={\rm Z}_\odot$.

Our simulations show that the dynamics is essential to map the population of BH binaries and that of X-ray binaries in young SCs. Firstly, the rate of BH binaries formed through dynamical exchanges is very high for all the considered metallicities. Secondly, the hardening of the binary due to three-body encounters is an important mechanism to induce RLO in binaries hosting BHs.

 The role of dynamics is particularly important for MSBHs.
In SCs with $Z=0.01$ and 0.1 Z$_\odot{}$, about 75 per cent of simulated MSBHs form from single stars and become members of binaries through dynamical exchanges in the first 100 Myr of the SC life. This is a factor of $\gtrsim{}3$ more efficient than in the case of low-mass ($<25$ M$_\odot{}$) stellar BHs. We show that the vast majority of MSBHs in RLO binaries originated from single stars and went through a dynamical exchange (Table~3 and Section 4.3). This is consistent with the qualitative predictions from the theory of three-body encounters combined with stellar evolution (Section 4.6).

In fact, 
the higher is the mass of the BH involved, the more important the effects of dynamics on X-ray binaries, as the rate of three-body encounters scales approximately with the total mass of the binary (e.g. Sigurdsson \&{} Phinney 1993). Furthermore, the probability for a single BH to become a binary member through a dynamical exchange is higher if the mass of the single BH is higher than the mass of one of the two components of the binary (e.g. Hills \&{} Fullerton 1980).
  Therefore, MSBHs are even more influenced by dynamics than their low-mass analogues.

Our results agree with the basic conclusion by L10: it is very hard that MSBHs power RLO X-ray binaries, if they form in primordial binaries that evolve unperturbed. On the other hand, we find that the situation is completely different for MSBHs that form in dynamically active environments (such as dense young SCs). In these environments, MSBHs efficiently power  RLO X-ray binaries, as a consequence of dynamical exchanges and three-body encounters, which dramatically alter the orbital properties of primordial binaries. 
Our results are also in fair agreement with those by B06, although there are some important differences, as B06 study IMBHs born from the runaway collapse in denser SCs, do not account for metallicity-dependent stellar evolution, and have a much simplified dynamical treatment. Both our paper and B06 find that BHs with mass $>50$ M$_\odot{}$ spend a large fraction of the simulated time with a companion star and that these massive BHs can power RLO binaries as a consequence of three-body encounters.





Our preliminary results show that the study of MSBHs is very promising, but there is a lot of work still to be done. 
First, the sample of MSBHs obtained from our simulations is statistically small (112 and 118 MSBHs for the simulations with $Z=0.1$ and $=0.01$ Z$_\odot$, respectively). Although our main results are physically well motivated and in agreement with previous studies about three-body encounters (e.g., Hills \&{} Fullerton 1980; Sigurdsson \&{} Phinney 1993) and population synthesis models (e.g. L10), a much larger sample is required to study in detail the properties of MSBHs in RLO binaries.

From the theoretical point of view, different models of stellar evolution and especially of mass loss by stellar winds need to be investigated with the same approach. Different environments (not only dense young SCs, but also globular clusters and open clusters) need to be studied with larger statistics. Furthermore, in this paper we assume a primordial binary fraction $f_{\rm PB}=0.1$. Higher binary fractions are not unrealistic (e.g. Delgado-Donate et al. 2004; Ivanova et al. 2005; Sollima et al. 2010)  and must be considered in future simulations. The code described in this paper traces accurately the dynamics of close encounters, but adopts simplified recipes for stellar and binary evolution, as well as for MT. The MT systems individuated by our simulations will be re-simulated with more accurate binary evolution codes, to trace how they evolve, which are their expected emission features and whether they can explain a fraction of the ULXs.

A fundamental question is where we can search for MSBHs. MW globular clusters can reach metallicities as low as $Z=0.01\,{}{\rm Z}_\odot$. Thus, they may host MSBHs, single or with low-mass companions. Metallicities of the order of $\sim{}0.1-0.2\,{}{\rm Z}_\odot$ are not infrequent in nearby irregular galaxies (e.g. IC~10), which are good candidates to further investigate the metallicity$-$BH mass connection.




\section*{Acknowledgments}
We thank the anonymous referee for her/his critical reading of the manuscript, which improved it. 
We made use of the public software package Starlab (version 4.4.4) and of the SAPPORO library (Gaburov, Harfst \&{} Portegies Zwart 2009) to run Starlab on graphics processing units (GPUs). We acknowledge all the developers of Starlab, and especially its primary authors: Piet Hut, Steve McMillan, Jun Makino, and Simon Portegies Zwart. We thank the authors of SAPPORO, and in particular E. Gaburov, S. Harfst and S.  Portegies Zwart. We also thank Paola Marigo, Alessia Gualandris, Antonella Vallenari, Rosanna Sordo, Laura Greggio, Mauro Barbieri and Monica Colpi for stimulating discussions.
 We acknowledge the CINECA Award N.  HP10CXB7O8 and HP10C894X7, 2011 for the availability of high performance computing resources and support. MM, LZ and ER acknowledge financial support from INAF through grant PRIN-2011-1. LZ acknowledges financial support from ASI/INAF grant no. I/009/10/0. AB acknowledges financial support from MIUR through grant PRIN-2009-1. MM thanks the Aspen Center for Physics, where part of this work was done.

\appendix
\section{The importance of primordial binaries}
An initial binary fraction $f_{\rm PB}=0.1$ was adopted for all the simulations presented in the main text. In this Appendix, we show the results of a set of 100 supplementary runs without primordial binaries ($f_{\rm PB}=0$). While these simulations are quite unrealistic, as primordial binaries are expected to exist in SCs, they are useful to illustrate the importance of primordial binaries. All the other properties of the SCs are the same as in the main text. In order to better see the effects on MSBHs, we maximize their formation probability by taking  $Z=0.01$ Z$_\odot{}$.

\begin{table*}
\begin{center}
\caption{Statistics of the simulated BHs, if $f_{\rm PB}=0$.} \leavevmode
\begin{tabular}[!h]{llllllllllll}
\hline
$Z$ [Z$_{\odot}$]
& $N_{\rm BH, cl}$
& $f_{\rm bin}$
& $f_{\rm sin}$
& $N_{\rm exch}$
& $t_{\rm life}$ [Myr]
& $f_{\rm W}$
& $f_{\rm W,\,{}exch}$
& $f_{\rm RL}$
& $f_{\rm RL,\,{}exch}$
& $f_{\rm MSBH}$
& $f_{\rm RL,\,{}MSBH}$
\\
\hline
0.01 & 8.12   & 0.27 & 0.27 & 2.5 & 38.5 & 0.034 & 0.034 & 0.009 & 0.009 & 0.17  &  0.43 \\
\noalign{\vspace{0.1cm}}
\hline
\end{tabular}
\begin{flushleft}
\footnotesize{The quantities shown in this Table are the same as defined in columns $1-10$ and $13-14$ of Table~2.}
\end{flushleft}
\end{center}
\end{table*}

\begin{figure}
\center{{
\epsfig{figure=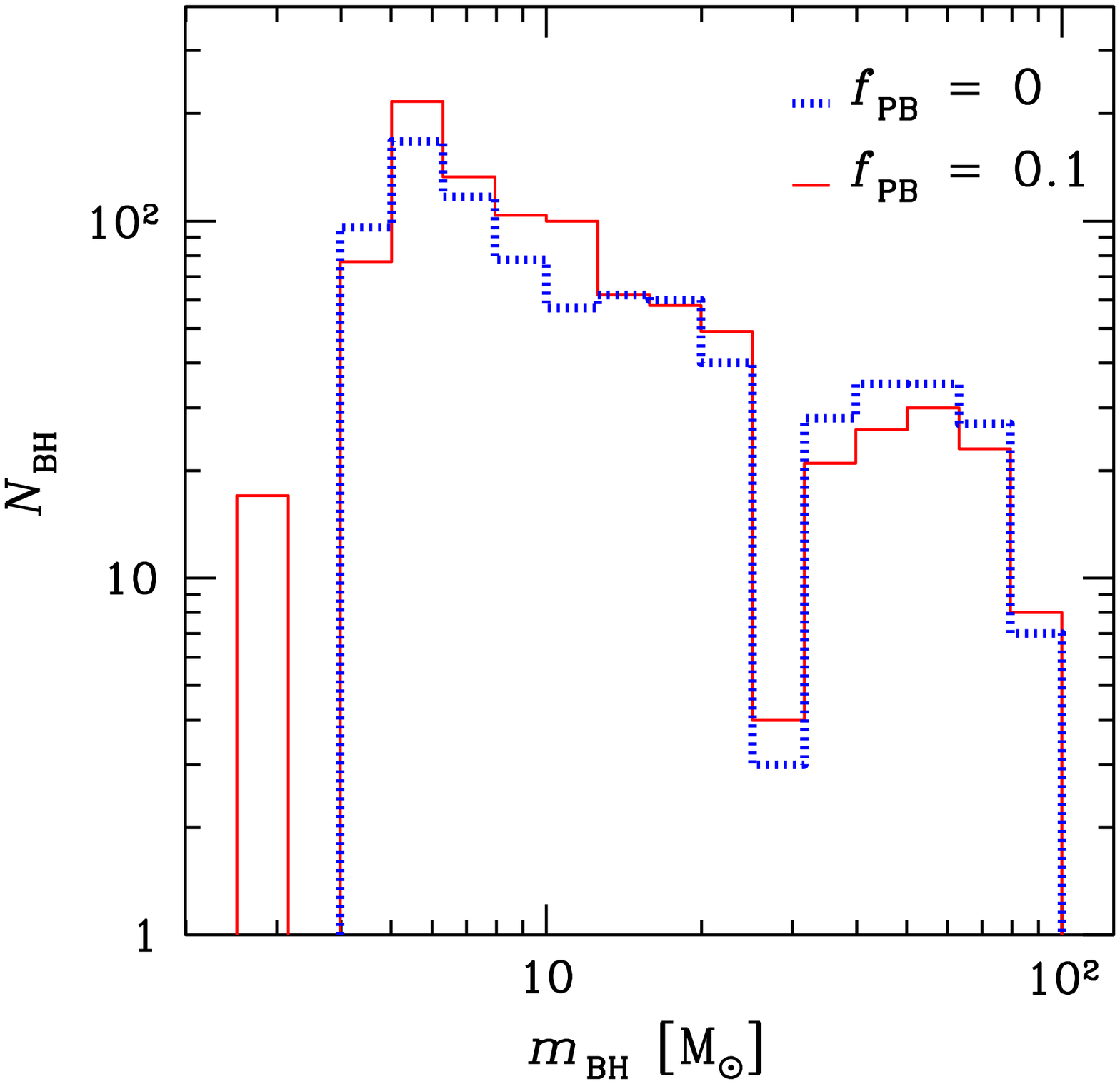,height=6.5cm} 
}}
\caption{\label{fig:figA1}
Mass distribution of BHs in the simulations with $Z=0.01$ Z$_\odot{}$, including a fraction of binaries $f_{\rm PB}=0$ (dotted line, blue on the web) and $f_{\rm PB}=0.1$ (solid line, red on the web). BH masses are calculated at the time of formation of the BHs (i.e. do not account for later mergers and/or accretion). The solid-line histogram is the same as in Fig.~\ref{fig:fig3}.}
\end{figure}

Table~A1 shows some of the most relevant properties of the BH population in the runs with $f_{\rm PB}=0$ and $Z=0.01$ Z$_\odot{}$. Some interesting considerations can be drawn from the comparison of Table~A1 with Table~2. First, the number of BHs per cluster is  about 10 per cent lower than in the case with  $f_{\rm PB}=0.1$. This is because a number of BHs form from the merger of relatively small stars (13-25 M$_\odot$) in primordial binaries, if $f_{\rm PB}>0$. This is evident from Fig.~\ref{fig:figA1}, where the initial mass spectrum of BHs in the runs with $f_{\rm PB}=0.1$ and with $f_{\rm PB}=0$ are compared. No BHs with mass $<4$ M$_\odot{}$ form in the case with $f_{\rm PB}=0$, whereas all the BHs with mass $\sim{}3$ M$_\odot{}$ formed in the runs with $f_{\rm PB}=0.1$ are remnants of merged progenitors. We stress that the mass of BHs born from merged progenitors depends strongly on the assumptions about mass loss during the merger.

The fraction of BHs that become members of a binary at least once in the simulation ($f_{\rm bin}$) in Table~A1 is almost the same as in Table~2. On the other hand, $f_{\rm bin}=f_{\rm sin}$ by construction in the case with $f_{\rm PB}=0$. Thus, the number of exchanges that produce BH binaries is higher if $f_{\rm PB}=0$. Similarly, the average number of exchanges per binary ($N_{\rm exch}$) is almost double in the case of $f_{\rm PB}=0$ than in the case of $f_{\rm PB}=0.1$. These results indicate that exchanges involving BHs are more numerous if $f_{\rm PB}=0$. 

\begin{figure*}
\center{{
\epsfig{figure=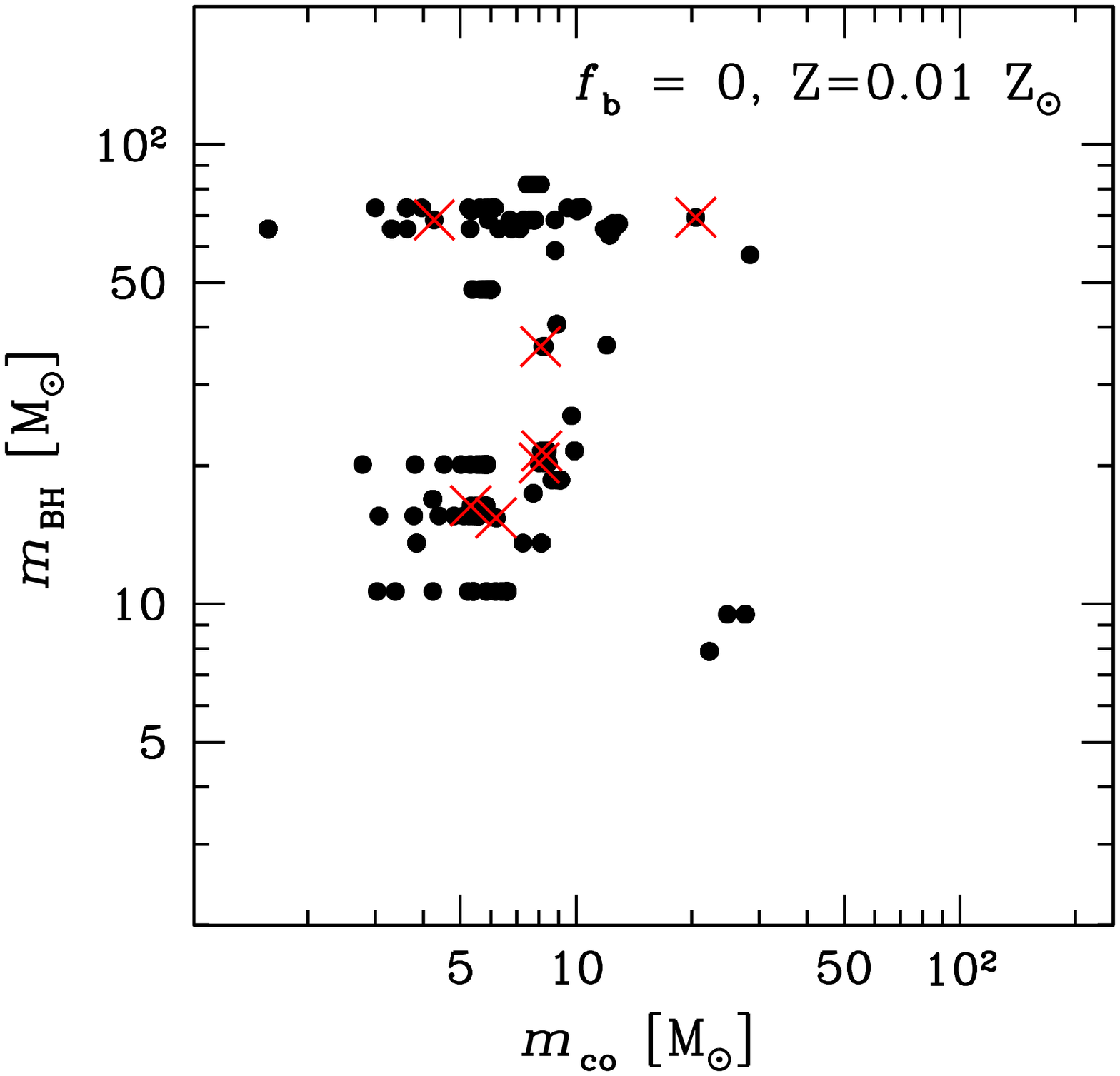,height=5.5cm} 
\epsfig{figure=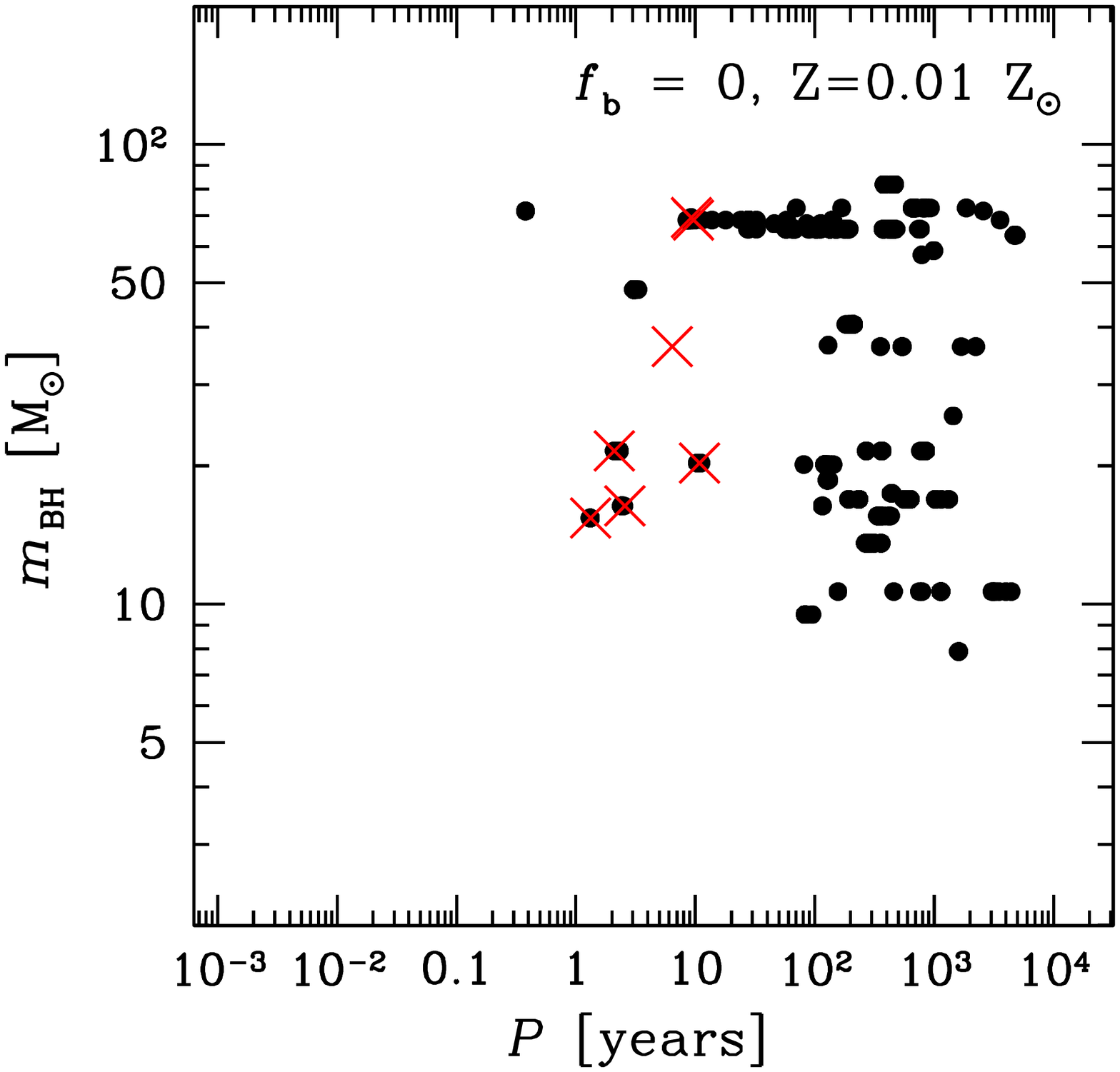,height=5.5cm}
\epsfig{figure=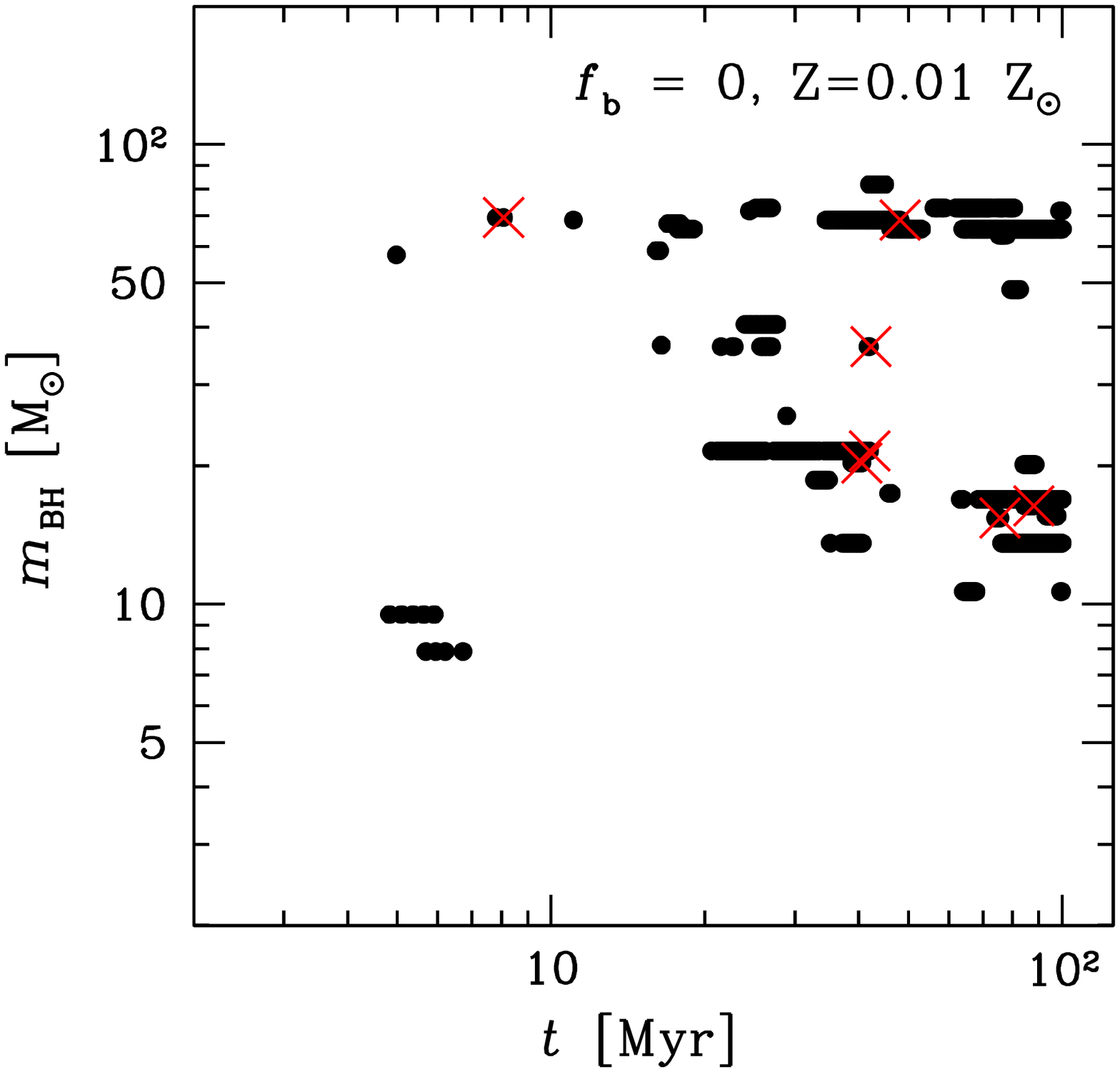,height=5.5cm}  
}}
\caption{\label{fig:figA2}
Mass of the BH versus mass of the companion star (left-hand panel), versus period (central panel) and versus the time elapsed since the beginning of the simulation (right-hand panel) for the 100 runs with $Z=0.01$ Z$_\odot$ and $f_{\rm PB}=0$. Filled circles: wind-accretion systems; crosses (red on the web): RLO systems (at the first RLO epoch). Each system can be identified by more than one point, when the mass of the secondary changes significantly (because of mass losses or because of dynamical exchange), when the period changes significantly and when the accretion lasts for more than one snapshot in the left-hand, central and right-hand panel, respectively. A cross and a circle almost superimposed indicate that the same system passes from wind-accreting to RLO (or {\it vice versa}).}
\end{figure*}

Furthermore, the properties of accreting BH binaries are considerably affected by the assumed fraction of primordial binaries. Wind-accreting systems and especially  RLO systems are strongly suppressed if $f_{\rm PB}=0$. This can be easily understood in light of the formation mechanism of binaries. If primordial binaries are allowed to form, then a number of primordial binary systems are expected to be sufficiently close to start RLO at early times, as a consequence of stellar evolution. In the absence of primordial binaries, binaries can form only through  the dynamical interaction of three single stars (hereafter, three-body capture) and through a tidal interaction between two single stars (hereafter, tidal capture). Three-body capture binaries are generally wide and  eccentric (Hut et al. 1992), whereas tidal-capture binaries are very hard and generally merge (e.g., Portegies Zwart et al. 1997). These two mechanisms are unlikely for most of the SC life, but are enhanced during core collapse (Spitzer 1987). 
In our simulations, we do not include any recipes for tidal-capture binaries. Thus we have only three-body capture binaries. Only a small fraction ($f_{\rm RL}/f_{\rm bin}\sim{}0.03$) of BH binaries born from three-body capture survives ionization and becomes hard enough (through three-body encounters) to start RLO within 100 Myr since the beginning of the simulation. 
The right-hand panel of Fig.~\ref{fig:figA2} confirms this consideration, by showing that all RLO systems switch on at relatively late times.

\begin{table*}
\begin{center}
\caption{Statistics of the simulated MSBHs, if $f_{\rm PB}=0$.} \leavevmode
\begin{tabular}[!h]{llllllllll}
\hline
$Z$ [Z$_{\odot}$]
& $N_{\rm MSBH, cl}$
& $f^{\rm MSBH}_{\rm bin}$
& $f^{\rm MSBH}_{\rm sin}$
& $N^{\rm MSBH}_{\rm exch}$
& $t^{\rm MSBH}_{\rm life}$ [Myr]
& $f^{\rm MSBH}_{\rm W}$
& $f^{\rm MSBH}_{\rm W,\,{}exch}$
& $f^{\rm MSBH}_{\rm RL}$
& $f^{\rm MSBH}_{\rm RL,\,{}exch}$
\\
\hline
0.01 & 1.36   & 0.85 & 0.85 & 2.6 & 51.8 & 0.110 & 0.110 & 0.022 & 0.022  \\
\noalign{\vspace{0.1cm}}
\hline
\end{tabular}
\begin{flushleft}
\footnotesize{The quantities shown in this Table are the same as defined in columns $1-10$ of Table~3.}
\end{flushleft}
\end{center}
\end{table*}
The last two columns of Table~A1 suggest another important consideration: if $f_{\rm PB}=0$, MSBHs are much more efficient than low-mass BHs in powering RLO systems. In fact, the fraction of RLO systems powered by MSBHs ($f_{\rm RL,\,{}MSBH}=0.43$) is about twice as high as the fraction of MSBHs with respect to all BHs ($f_{\rm MSBH}=0.17$). The bias towards high BH masses in RLO systems appears even more evident if we look at Fig.~\ref{fig:figA2}: all the BHs powering RLO systems have masses $\ge{}15$ M$_\odot{}$. This is a strong confirmation that the more a BH is massive, the more its chances of entering a binary by dynamical exchange are high.

Finally, Table~A2 shows some  of the most relevant properties of MSBHs in the runs with $f_{\rm PB}=0$. From the comparison with Table~3, it is apparent that more MSBHs can form in the absence of primordial binaries ($N_{\rm MSBH, cl}=1.36$ and 1.18 if $f_{\rm PB}=0$ and 0.1, respectively). If $f_{\rm PB}>0$, the mass loss following the CE phase in a primordial binary tends to produce smaller BH masses.
Finally, the fraction of RLO systems powered by MSBHs is a factor of 10 smaller if $f_{\rm PB}=0$ than if $f_{\rm PB}=0.1$. 

In summary, all the aspects discussed in this Appendix indicate that primordial binaries have a crucial importance for the population of X-ray binaries in SCs.
\section{Different stellar evolution and SN explosion recipes}
\begin{table*}
\begin{center}
\caption{Statistics of the simulated BHs, if the maximum mass of BHs is assumed to be $m_{\rm max}=25$ M$_\odot{}$.} \leavevmode
\begin{tabular}[!h]{llllllllllll}
\hline
$Z$ [Z$_{\odot}$]
& $N_{\rm BH, cl}$
& $f_{\rm bin}$
& $f_{\rm sin}$
& $N_{\rm exch}$
& $t_{\rm life}$ [Myr]
& $f_{\rm W}$
& $f_{\rm W,\,{}exch}$
& $f_{\rm RL}$
& $f_{\rm RL,\,{}exch}$
& $f_{\rm MSBH}$
& $f_{\rm RL,\,{}MSBH}$
\\
\hline
0.01 & 9.58   & 0.29 & 0.19 & 1.1 & 30.9 & 0.050 & 0.017 & 0.050 & 0.008 & 0.14  &  0.04 \\
\noalign{\vspace{0.1cm}}
\hline
\end{tabular}
\begin{flushleft}
\footnotesize{The quantities shown in this Table are the same as defined in columns $1-10$ and $13-14$ of Table~2.}
\end{flushleft}
\end{center}
\end{table*}
\begin{table*}
\begin{center}
\caption{Statistics of the simulated MSBHs,  if the maximum mass of BHs is assumed to be $m_{\rm max}=25$ M$_\odot{}$.} \leavevmode
\begin{tabular}[!h]{llllllllll}
\hline
$Z$ [Z$_{\odot}$]
& $N_{\rm MSBH, cl}$
& $f^{\rm MSBH}_{\rm bin}$
& $f^{\rm MSBH}_{\rm sin}$
& $N^{\rm MSBH}_{\rm exch}$
& $t^{\rm MSBH}_{\rm life}$ [Myr]
& $f^{\rm MSBH}_{\rm W}$
& $f^{\rm MSBH}_{\rm W,\,{}exch}$
& $f^{\rm MSBH}_{\rm RL}$
& $f^{\rm MSBH}_{\rm RL,\,{}exch}$
\\
\hline
0.01 & 1.34   & 0.69 & 0.61 & 1.5 & 37.3 & 0.030 & 0.030 & 0.015 & 0.015 \\
\noalign{\vspace{0.1cm}}
\hline
\end{tabular}
\begin{flushleft}
\footnotesize{The quantities shown in this Table are the same as defined in columns $1-10$ of Table~3.}
\end{flushleft}
\end{center}
\end{table*}
In this paper, we follow the stellar evolution recipes described in Section 3.1 and we assume that stars with $m_{\rm fin}\ge{}40$ M$_\odot{}$ directly collapse to MSBHs. Different stellar evolution recipes and different assumptions for the end of massive star life have important effects on the presented results. A complete comparison between different stellar evolution recipes will be done in the next papers of the series. In this Appendix, we just show a simplified, rather extreme case, in which the formation of MSBHs is strongly suppressed. In particular, we assume that the maximum BH mass is $m_{\rm max}=25$ M$_\odot{}$, even for $Z=0.01$ Z$_\odot{}$. Thus, in these runs the MSBHs can have only $m_{\rm BH}=25$ M$_\odot{}$.

We performed 50 runs of SC evolution assuming $m_{\rm max}=25$ M$_\odot{}$, $Z=0.01$ Z$_\odot{}$ and leaving all the other properties of the SCs as described in Section 3. The main results for the BH population and for X-ray binaries powered by BHs are summarized in Tables~B1 and B2. Table~B1 shows that  $N_{\rm BH, cl}$, $f_{\rm bin}$, $f_{\rm sin}$ and $N_{\rm exch}$ are very similar to the case with $Z=0.01$ Z$_\odot{}$ described in Table~2. The most significant difference between Table~B1 and Table~2 is the value of $f_{\rm RL,\,{}MSBH}$. The fraction of MSBHs that power RLO systems is much smaller in Table~B1 than in Table~2 (0.04 versus 0.24). This confirms that MSBHs in the mass range $25-80$ M$_\odot{}$ are much more efficient in powering RLO systems than BHs with $m_{\rm max}=25$ M$_\odot{}$.
The same conclusion can be derived from the comparison between Table~3 (for $Z=0.01$ Z$_\odot{}$) and Table~B2, where the statistics for MSBHs is shown. The MSBHs in Table~B2 (whose mass is $25$ M$_\odot{}$ by construction) are members of binaries and power RLO systems less often than the MSBHs in Table~3.
\end{document}